\documentclass[a4paper,fleqn]{cas-dc}

\usepackage[numbers, sort&compress]{natbib}
\def\tsc#1{\csdef{#1}{\textsc{\lowercase{#1}}\xspace}}
\tsc{WGM}
\tsc{QE}
\usepackage{url}
\usepackage{amsmath,amssymb,amsfonts}
\usepackage{algorithmic}
\usepackage{graphicx}
\usepackage{textcomp}
\usepackage{bm,multirow, booktabs}
\usepackage{float}
\usepackage[ruled]{algorithm2e}

\usepackage[colorlinks]{hyperref} 
\definecolor{revisecolor}{HTML}{0F35A6}

\usepackage{marginnote}

\def\tX{\mathcal{X}}
\def\mX{\mathbf{X}}
\def\mY{\mathbf{Y}}
\def\vx{\mathbf{x}}
\def\tmul{*_\mathsf{T}}  
\def\t{\mathcal}

\newtheorem{mydef}{Definition}
\newtheorem{mythrm}{Theorem}
\newtheorem{mylem}{Lemma}
\begin{document}
\let\WriteBookmarks\relax
\def\floatpagepagefraction{1}
\def\textpagefraction{.001}

\shorttitle{T$^2$LR-Net}    

\shortauthors{Y. Zhang \textit{et~al.}}  

\title[mode=title]{T$^2$LR-Net: An Unrolling Network Learning Transformed Tensor Low-Rank Prior for Dynamic MR Image Reconstruction}%

\author[1]{Yinghao Zhang}[orcid=0000-0001-8501-6180]
\author[1]{Peng Li}[orcid=0000-0001-5015-4364]
\author[1]{Yue Hu}[orcid=0000-0002-4648-611X]
\cormark[1]
\cortext[cor1]{Corresponding author: e-mail: huyue@hit.edu.cn}
\address[1]{School of Electronics and Information Engineering, Harbin Institute of Technology, Harbin, China}

\begin{abstract}
  The tensor low-rank prior has attracted considerable attention in dynamic MR reconstruction. Tensor low-rank methods preserve the inherent high-dimensional structure of data, allowing for improved extraction and utilization of intrinsic low-rank characteristics. However, most current methods are still confined to utilizing low-rank structures either in the image domain or predefined transformed domains. Designing an optimal transformation adaptable to dynamic MRI reconstruction through manual efforts is inherently challenging. In this paper, we propose a deep unrolling network that utilizes the convolutional neural network (CNN) to adaptively learn the transformed domain for leveraging tensor low-rank priors. Under the supervised mechanism, the learning of the tensor low-rank domain is directly guided by the reconstruction accuracy. Specifically, we generalize the traditional t-SVD to a transformed version based on arbitrary high-dimensional unitary transformations and introduce a novel unitary transformed tensor nuclear norm (UTNN). Subsequently, we present a dynamic MRI reconstruction model based on UTNN and devise an efficient iterative optimization algorithm using ADMM, which is finally unfolded into the proposed T$^2$LR-Net. Experiments on two dynamic cardiac MRI datasets demonstrate that T$^2$LR-Net outperforms the state-of-the-art optimization-based and unrolling network-based methods.
\end{abstract}

% \begin{highlights}
%   \item Proposing a novel transformed t-SVD framework based on arbitrary unitary transform and developing the corresponding unitary transformed tensor nuclear norm (UTNN). 
%   \item Introducing a novel dynamic MR reconstruction model regularized by UTNN and developing an efficient UTNN-based iterative optimization algorithm using ADMM.
%   \item Establishing the T$^2$LR-Net by unrolling the iterations of the UTNN-based iterative optimization algorithm. 
%   \item Incorporating convolutional neural network to learn an optimal transform to explore the tensor low-rank prior in the transformed domain. 
%   \item Experiments on two cardiac cine MR datasets demonstrate that T$^2$LR-Net outperforms SOTA optimization-based and unrolling network-based methods.
%   \end{highlights}


\maketitle

\section{Introduction}
\label{sec:intro}

Dynamic magnetic resonance (MR) imaging is critical in several clinical applications, such as cardiac, perfusion, and vocal tract imaging. It captures more information than static MR imaging, resulting in a spatiotemporal three-dimensional MR data structure that aids in the identification of certain diseases, such as cardiovascular diseases. However, obtaining dynamic MR images with high spatiotemporal resolution within clinically acceptable scan time is still very challenging.

In clinical practice, imaging acceleration is achieved through undersampling k-space (Fourier domain). This compromises the Nyquist sampling theorem, leading to aliasing in the image domain. Therefore, effective algorithms are required for the reconstruction of undersampled data. Currently, sparsity in the wavelet \cite{lustig2007sparse}, temporal Fourier \cite{jung2009k}, and total variation domain \cite{ref_ktslr} has been employed for reconstruction, yielding satisfactory results. Additionally, due to the slow variations in the same organ, dynamic MR frames are, in fact, temporally correlated throughout the entire image sequence. Methods based on the low-rank Casorati matrix \cite{liang2007spatiotemporal} can effectively utilize this temporal redundancy information for reconstruction. Furthermore, approaches combining sparsity and low rank have been proposed, demonstrating superior reconstruction performance compared to single-constraint methods. These methods primarily fall into two categories. One is methods employing joint low-rank and sparse constraints, such as k-t SLR \cite{ref_ktslr}. The other involves separating low-rank and sparse components \cite{ref_matrix2}, decomposing dynamic MR images into a slowly varying background and a foreground part that is both small in proportion and exhibits noticeable motion. The mentioned methods above fall under the category of iterative optimization algorithms. They utilize algorithms such as ISTA \cite{beck2009fast} and ADMM \cite{boyd2011distributed} to solve the reconstruction model. However, these methods require manual tuning of hyperparameters and exhibit a relatively slow convergence speed.

Deep unrolling networks \cite{LeCun2010unroll} have recently shown promising results in dynamic MR reconstruction \cite{aggarwal2018modl,qin2018convolutional, huang2021deep}. These networks implement iterative optimization algorithms within the supervised framework of deep learning, eliminating the need for manual hyperparameter tuning. This approach not only leverages physics-driven priors but also harnesses the powerful learning capabilities of neural networks. ISTA-Net \cite{ref_ISTANET} employs the convolutional neural network (CNN) to develop a transformed sparse prior for reconstruction, while SLR-Net \cite{ref_slrnet} jointly utilizes the sparsity in the CNN transformed domain and the Casorati matrix low-rank prior. DCCNN \cite{ref_DCCNN} uses CNN to extract and leverage implicit deep image information for reconstruction. These methods have all achieved reconstruction performance surpassing those of traditional iterative optimization algorithms.
 
In recent years, the tensor low-rank prior \cite{ref_tensor} has emerged as a promising approach for accelerating dynamic MR imaging due to its ability to capture the inherent high-dimensional structure of the dynamic MR data. Nevertheless, unlike the unique definition of the matrix rank, the tensor rank has different definitions under the Candecomp/Parafac (CP) and Tucker decompositions. Yaman \emph{et al.} \cite{yaman2019low} successfully applied the low-CP-rank approximation for cardiac MR image reconstruction. However, determining the tensor CP rank is computationally expensive and NP-hard. In contrast, the Tucker rank has been employed by formulating a reconstruction model regularized by the sum of the nuclear norms (SNN) of the unfolding matrices \cite{ref_wmnn4}. The SNN and sparsity were incorporated in \cite{ref_wmnn2,liu2022highly} to further improve the reconstruction performance. However, the SNN is not the exact convex envelope of the sum of the Tucker ranks \cite{ref_snn3}, leading to suboptimal reconstruction performance at high acceleration factors. Some studies have attempted to address this limitation by leveraging the sparsity constraint on the core tensor \cite{he2016accelerated,christodoulou2018magnetic}, but this approach can be computationally complex. Recently, multilayer sparsity in tensor decomposition \cite{xue2021multilayer, xue2022laplacian} has also been proposed to leverage the low-rank properties of tensors.

Compared to CP and Tucker, the tensor singular value decomposition (t-SVD) \cite{ref_tsvd} offers a simpler decomposition form that can be computed by the matrix SVDs of the frontal slices. The induced tensor nuclear norm (TNN) \cite{ref_tnn1} is the exact convex envelope of the corresponding tensor rank. Due to the mathematical soundness of t-SVD, it has been widely used in tensor completion \cite{zhang2016exact,ref_pre_ttnn,lu2016trpc,ref_tnn1} and hyperspectral image restoration \cite{ref_tnn2,ref_pre_ttnn} with excellent results. Specifically, t-SVD is based on Fast Fourier Transform (FFT), indicating that tensor low-rank methods under the t-SVD framework utilize low-rank representation information in the FFT domain, which, however, may be imprecise and limited. Therefore, finding a suitable transformed domain and utilizing tensor low-rank priors in that domain may lead to superior reconstruction results. F2TNN \cite{ref_pre_ttnn} employed the framelet transform on the spatial dimensions and the FFT on the temporal dimension. FTNN \cite{jiang2020framelet} adopted the temporal framelet transformation. 
S2NTNN \cite{luo2022self, luo2022hlrtf} utilized fully connected layers in an unsupervised framework to approximate the required tensor low-rank transform domain. More generally, t-SVD can be extended to one-dimensional invertible \cite{ref_ttsvd} or unitary \cite{ref_ttnn} transformed versions to provide theoretical support for tensor low-rank priors in the corresponding transformed domain. 

However, several challenges remain to be addressed. Firstly, most of the existing works related to transformed tensor low-rank priors are concentrated on tensor completion. Further research is needed to explore their applicability and effectiveness in dynamic MR reconstruction tasks. The data acquisition in MRI occurs in the frequency domain, while the reconstruction target is an image. This cross-domain characteristic fundamentally distinguishes it from tensor completion tasks. Secondly, current research only focused on one-dimensional transformations, possibly due to considerations of the complexity of designing artificial transforms. However, this lack of flexibility in the design of the transform may result in suboptimal reconstruction performance. Additionally, the choice of the transform is crucial. Existing models require handcrafted and predefined transformations. However, there is no direct relationship between the transformation and the final reconstruction accuracy, making it challenging to find an appropriate transformation.

In this paper, we propose an unrolling network learning Transformed Tensor Low-Rank prior for dynamic MR imaging, termed T$^2$LR-Net. Specifically, We employ CNN to adaptively learn the tensor low-rank transformed domain from the dynamic MR dataset, aiming to maximize the reconstruction accuracy. Due to the supervised mechanism of the deep unrolling network, the reconstruction error serves as a loss function that directly guides the selection of the transformed domain. Furthermore, given the high-dimensional transformations of CNNs, the existing one-dimensional transformation-based t-SVD framework is no longer applicable. Therefore, we extend t-SVD to arbitrary multi-dimensional unitary transformations, providing theoretical support for the proposed T$^2$LR-Net. Our contributions can be summarized as follows:
\begin{itemize}
  \item We propose a novel transformed t-SVD framework based on \emph{arbitrary} unitary transform, which can be multi-dimensional. This distinguishes our approach from existing t-SVDs that predominantly rely on one-dimensional transforms. The corresponding unitary transformed tensor nuclear norm (UTNN) is also developed and proved to be the convex envelope of the transformed tensor sum rank.
  \item We introduce a novel dynamic MR reconstruction model regularized by UTNN and develop an efficient UTNN-based iterative optimization algorithm using ADMM. All subproblems of ADMM can be solved analytically, according to the mathematical derivation of the proposed UTNN.
  \item We establish the T$^2$LR-Net by unrolling the iterations of the UTNN-based iterative optimization algorithm into multiple iteration modules. In each module, the convolutional neural network (CNN) is used to learn an independent transformation to explore the specific intrinsic low-rank property of the data. The supervised mechanism in the deep unrolling network directly guides the learning of the transformed domain, leading to a significant improvement in reconstruction performance.
\end{itemize}

A preliminary version of this work was presented in IEEE ISBI 2023 \cite{zhang2023tlr}. However, this journal paper has undergone significant changes. We have further provided the mathematical derivations for UTNN and the tensor singular value thresholding method. Experiments involving multi-coil and prospective scenarios have been validated. Additionally, in the discussion section, we performed a comprehensive analysis of the interplay between CNN and tensor low-rank prior. It is worth noting that the authors of \cite{zhang2020video} also propose leveraging the tensor low-rank properties in the CNN-transform domain. However, they focus on video synthesis, whereas our work targets dynamic MR reconstruction. The differences arise from the inherent Fourier domain sampling characteristics of MRI, distinguishing it from video applications. In their paper, the t-SVD architecture fundamentally relies on one-dimensional transformations \cite{ref_ttsvd2}, yet they directly generalize this using a three-dimensional transformation like 2D CNN, lacking theoretical assurance. In contrast, our work extends t-SVD to multidimensional transformations, covering both one-dimensional and three-dimensional transformations. We provide detailed proofs in the \hyperlink{appendix}{Appendix}, thus offering robust mathematical theoretical support. Additionally, their method utilized the ISTA algorithm, while we design an efficient ADMM algorithm to solve the proposed reconstruction model.

The rest of this paper is organized as follows. Section \ref{background} introduces the preliminaries and the generic dynamic MR image reconstruction models. Section \ref{sec:methodology} describes the proposed method, and the experiments and results are shown in Section \ref{sec:experiments}. Discussion and conclusion are provided in Section \ref{discuss} and \ref{conclude}, respectively.

\section{Background}
\label{background}

\subsection{Notations and Preliminaries}
In this paper, we denote tensors by Euler script letters, e.g., $\tX$, matrices by bold capital letters, e.g., $\mX$, vectors by bold lowercase letters, e.g., $\vx$, and scalars by lowercase letters, e.g., $x$. For a 3-way tensor $\tX \in \mathbb{C}^{n_1 \times n_2 \times n_3}$, we denote $\tX^{(i)}$ as the $i$-th frontal slice, $\tX(:,:,i),i=1,2,...,n_3$. Operators or transformations are denoted by sans serif capital letters, e.g., $\mathsf{T}$, $\mathsf{F}$.

\begin{mylem}[Unitary transform]
  
A transform $\mathsf{T}$ is the unitary transform only if it preserves the Frobenius norm and inner product \cite{horn2012matrix}, i.e., 
\begin{equation}
    \|\tX\|_F = \|\hat{\tX}_\mathsf{T}\|_F \ and \  <\tX, \mathcal{B}> = <\hat{\tX}_\mathsf{T}, \hat{\mathcal{B}}_\mathsf{T}>.
\end{equation}
where $\tX \in \mathbb{C}^{n_1 \times n_2 \times n_3}$, $\mathcal{B} \in \mathbb{C}^{n_2 \times n_4 \times n_3}$, $\mathsf{T}:\mathbb{C}^{n_1 \times n_2 \times n_3} \rightarrow \mathbb{C}^{m_1 \times m_2 \times m_3}$, and $\hat{\tX}_\mathsf{T} = \mathsf{T}(\tX)$.
\end{mylem}

The Frobenius norm of 3-way tensor is defined by $\|\tX\|_F = \sqrt {\sum_{ijk}|a_{ijk}|^2}$. The inner product between $\tX$ and $\t{B}$ is calculated by the sum of inner products of all frontal slices, i.e., $<\tX, \mathcal{B}> = \sum_{i=1}^{n_3}<\tX^{(i)}, \mathcal{B}^{(i)}>$, and for two matrices $\mX$ and $\mY$, $<\mX, \mY> = \operatorname{trace}(\mX^H \mY)$. Due to the unitary property, $\tX$ can be also obtained by applying the Hermitian transpose transform $\mathsf{T}^H: \mathbb{C}^{m_1 \times m_2 \times m_3} \rightarrow \mathbb{C}^{n_1 \times n_2 \times n_3}$ on $\hat{\tX}_\mathsf{T}$, i.e., $\tX = \mathsf{T}^H(\hat{\tX}_\mathsf{T})$.

The block diagonal matrix based on the frontal slices of $\hat{\tX}_\mathsf{T}$ is denoted as follows,
\begin{equation}
  \footnotesize
  \label{bdiag}
  \overline{\tX_\mathsf{T}} = 
  \left(
      \begin{array}{llll}
          \hat{\tX}^{(1)}_\mathsf{T} & & & \\
          & \hat{\tX}^{(2)}_\mathsf{T} & & \\
          & & \ddots & \\
          & & & \hat{\tX}^{(m_3)}_\mathsf{T}
      \end{array}
  \right),
\end{equation}
which can be converted back into a tensor by the following fold operator,
\begin{equation}
  \operatorname{fold}(\overline{\tX_\mathsf{T}}) = \hat{\tX}_\mathsf{T}.
\end{equation}
Thus, we can obtain the following relationship,
\begin{equation}
  \label{Fnorm}
  \|\tX\|_F = \|\hat{\tX}_\mathsf{T}\|_F = \|\overline{\tX_\mathsf{T}}\|_F,
\end{equation}
\begin{equation}
  \label{inner_prod}
    <\tX, \mathcal{B}> = <\tX_\mathsf{T}, \mathcal{B}_\mathsf{T}>
    = <\overline{\tX_\mathsf{T}}, \overline{\t{B}_\mathsf{T}}>.
\end{equation}

\begin{mydef}[$\mathsf{T}$-product]
  The $\mathsf{T}$-product of $\tX \in \mathbb{C}^{n_1 \times n_2 \times n_3}$ and $\mathcal{B} \in \mathbb{C}^{n_2 \times n_4 \times n_3}$ based on a unitary transform $\mathsf{T}$ is a tensor $\mathcal{C} \in \mathbb{C}^{n_1 \times n_4 \times n_3}$, which can be expressed as 
  \begin{equation}
      \label{eq:tmul}
      \mathcal{C} = \tX *_\mathsf{T} \mathcal{B} = \mathsf{T}^H \circ \operatorname{fold}(\overline{\tX_\mathsf{T}} \times \overline{\t{B}_\mathsf{T}}),
  \end{equation}
  where `$\times$' denotes the standard matrix product and $\circ$ is the composition operator. 
\end{mydef}

Note that the $\mathsf{T}$-product can be expressed as:
\begin{equation}
  \label{tprod_2}
  \overline{\t{C}_\mathsf{T}} = \overline{\tX_\mathsf{T}} \times \overline{\t{B}_\mathsf{T}},
\end{equation}
which means that the $\mathsf{T}$-product, in the transformed domain, is conducted by slice-wise matrix multiplication of each frontal slice of $\hat{\tX}_\mathsf{T}$ and $\hat{\mathcal{B}}_\mathsf{T}$  \cite{ref_ttnn,ref_ttsvd}. 

Given $\tX \in \mathbb{C}^{n_1 \times n_2 \times n_3}$, the identity tensor is defined by $(\mathcal{I}_\mathsf{T})_{n_1} *_\mathsf{T} \tX = \tX *_\mathsf{T} (\mathcal{I}_\mathsf{T})_{n_2} = \tX$, where $(\mathcal{I}_\mathsf{T})_{n_1} \in \mathbb{C}^{n_1 \times n_1 \times n_3}$ is the identity tensor with the first two dimensions equal to $n_1$.
The Hermitian transpose is denoted as $\tX^H = \mathsf{T}^H[\operatorname{fold}(\overline{\tX_\mathsf{T}}^H)] \in \mathbb{C}^{n_2 \times n_1 \times n_3}$, 
and the unitary tensor $\mathcal{Q} \in \mathbb{C}^{n \times n \times n_3}$ is denoted as
$\mathcal{Q}^H \tmul \mathcal{Q} = \mathcal{Q} \tmul \mathcal{Q}^H = \mathcal{I}_\mathsf{T}.$

\subsection{Problem Formulation}
\label{dmri model}

The data acquisition of dynamic MR imaging can be modeled as
\begin{equation}
  \label{dmri_acquisition}
\mathbf b = \mathsf A(\mathcal{X}) +\mathbf{n},
\end{equation}
where $\mathcal{X} \in \mathbb{C}^{n_{x} \times n_{y} \times n_{t}}$ denotes distortion-free dynamic MR data, $n_x$, $n_y$ denote the spatial coordinates, $n_t$ is the temporal coordinate, $\mathbf b \in \mathbb{C}^{m}$ is the observed undersampled $k$-space data, $\mathsf A: \mathbb{C}^{n_{x} \times n_{y} \times n_{t}} \rightarrow \mathbb{C}^m$ is the encoding operator, and $\mathbf{n} \in \mathbb{C}^{m}$ is the Gaussian distributed white noise. 

In the single-coil scenario, $\mathsf A$ is the Fourier sampling operator, i.e., $\mathsf A = \mathsf{F_u}$. In the multi-coil scenario, $\mathsf A = \mathsf{F_u} \circ \mathsf{C}$, where $\mathsf{C}$ denotes the coil sensitivity map and $\circ$ denotes the composition operator. Specifically in the Cartesian sampling cases, $\mathsf{F_u} = \mathsf{S} \circ \mathsf{F}$, where $\mathsf S: \mathbb{C}^{n_{x} \times n_{y} \times n_{t} \times n_{c}} \rightarrow \mathbb{C}^m$ denotes the sampling operator and $\mathsf{F}: \mathbb{C}^{n_{x} \times n_{y} \times n_{t} \times n_{c}} \rightarrow \mathbb{C}^{n_{x} \times n_{y} \times n_{t} \times n_{c}} (\mathsf{F}^H \circ \mathsf{F}(\mathcal{X}) = \mathsf{F} \circ \mathsf{F}^H(\mathcal{X}) = \mathcal{X})$ is the unitary two-dimensional spatial Fourier transform on $x$ and $y$ axes.

The dynamic MR image reconstruction model is commonly formulated as the following optimization problem:
\begin{equation}
  \tX^* = \arg \min_{\tX} \frac 12\|\mathsf A(\tX) - \mathbf{b}\|_2^2 + \lambda \mathsf{R} (\tX),
\end{equation}
where $\|\mathsf A(\tX) - \mathbf{b}\|_2^2$ is the data fidelity term that guarantees the consistency between the $k$-space of the reconstruction and the observation, $\mathsf{R} (\tX)$ is the regularization term, %that exploits the tensor priors to improve the reconstruction performance, 
and $\lambda$ is the balancing parameter.%, $\mathsf{R}: \mathbb{C}^{n_{1} \times n_{2} \times n_{3}} \rightarrow \mathbb{R}$ is the prior-extracting operator, e.g. MNN, TNN or CNN.

\section{Methodology}
\label{sec:methodology}

\subsection{Transformed t-SVD and UTNN Based on Arbitrary Unitary Transform}

In this part, we propose a novel transformed t-SVD framework based on arbitrary unitary transform. We introduce the unitary transformed tensor nuclear norm (UTNN) that corresponds to this framework. The UTNN is proved to be the convex envelope of the transformed tensor sum rank, providing a theoretical guarantee for utilizing UTNN to exploit the tensor low-rank prior.

\begin{mythrm}[Unitary Transformed t-SVD]
  
  The transformed t-SVD of $\tX \in \mathbb{C}^{n_1 \times n_2 \times n_3}$ based on arbitrary unitary transform $\mathsf{T}$ can be factorized as follows:
  \begin{equation}
    \label{ttsvd}
      \tX = \mathcal{U} \tmul \mathcal{S} \tmul \mathcal{V}^H,
  \end{equation}
  where $\t{U} \in \mathbb{C}^{n_1 \times n_1 \times n_3}$ and $\t{V}\in \mathbb{C}^{n_2 \times n_2 \times n_3}$ are unitary tensors with respect to $\mathsf{T}$-product, and $\t{S}$ is the core tensor of the transformed t-SVD.
\end{mythrm}
\textbf{\emph{proof.}} Given a three-dimensional tensor $\tX \in \mathbb{C}^{n_1 \times n_2 \times n_3}$, to calculate the transformed t-SVD based on arbitrary unitary transform $\mathsf{T}$, we first apply the unitary transformation and obtain transformed tensor $\hat{\tX}_\mathsf{T}$. Then, for each frontal slice of $\hat{\tX}_\mathsf{T}$, we compute the matrix SVD, 
\begin{equation}
  \hat{\tX}^{(i)}_\mathsf{T} = \hat{\t{U}}^{(i)}_\mathsf{T}\hat{\t{S}}^{(i)}_\mathsf{T}\hat{\t{V}}^{(i)H}_\mathsf{T},
\end{equation}
where $\hat{\t{U}}^{(i)}_\mathsf{T} \in \mathbb{C}^{n_1 \times n_1}$ and $\hat{\t{V}}^{(i)}_\mathsf{T} \in \mathbb{C}^{n_2 \times n_2}$ are unitary matrices, and $\hat{\t{S}}^{(i)}_\mathsf{T} \in \mathbb{C}^{n_1 \times n_2}$ is a diagonal matrix with the singular values of $\hat{\tX}^{(i)}_\mathsf{T}$ on the diagonal. Then, by rearranging all the unitary matrices and diagonal matrices into three block diagonal matrices, we can obtain $\overline{\t{U}_\mathsf{T}}$, $\overline{\t{V}_\mathsf{T}}$, and $\overline{\t{S}_\mathsf{T}}$, respectively. Therefore, the above equation can be rewritten as, 
\begin{equation}
  \label{eq_usv2}
  \overline{\mathcal{X}_\mathsf{T}} = \overline{\mathcal{U}_\mathsf{T}} \times \overline{\t{S}_\mathsf{T}} \times \overline{\t{V}_\mathsf{T}}^H.
\end{equation}
From equation \eqref{eq:tmul} and \eqref{tprod_2}, we can further convert this formula and obtain equation \eqref{ttsvd}. The unitary properties of $\t{U}$ and $\t{V}$ can be easily proved by verifying the definition of unitary tensor.

\begin{figure}[h]
  \centering
  \includegraphics[scale=0.3]{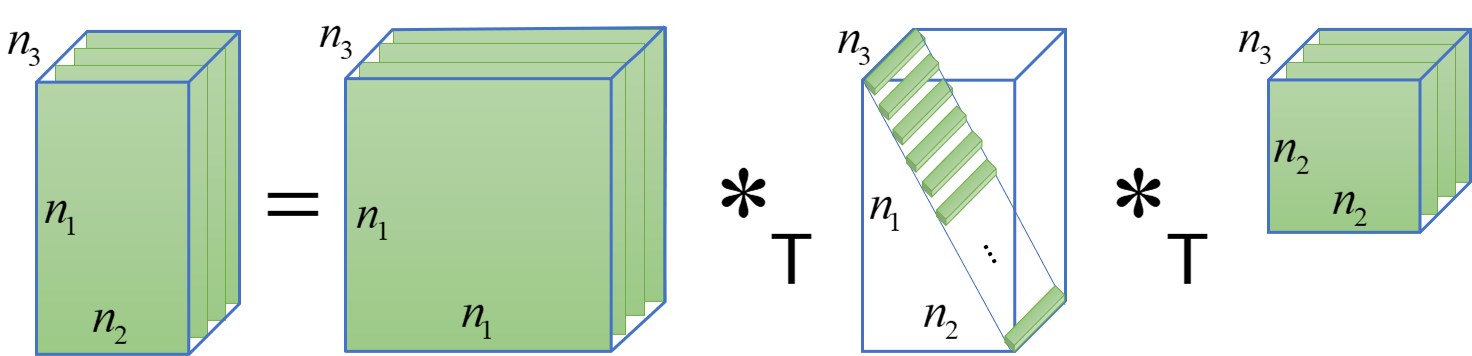}
  \caption{An illustration of the proposed transformed t-SVD factorization of a tensor with dimensions $n_1 \times n_2 \times n_3$.}
  \label{fig:ttSVD}
\end{figure}

An illustration of the proposed transformed t-SVD factorization is shown in Figure \ref{fig:ttSVD}, which has the similar structure as the other t-SVDs. Nevertheless, our main distinction from existing t-SVDs \cite{ref_tnn1,ref_ttnn,ref_ttsvd,ref_tsvd} is that current t-SVDs are established based on one-dimensional invertible or unitary transformations, while our transformation is an arbitrary unitary transformation that can be multidimensional. If we set the transformation to be one-dimensional, it can degrade to existing t-SVDs. This generalization aligns with our intention to leverage CNN for learning transformation because CNN operates as a multi-dimensional transformation. The t-SVDs based on one-dimensional transformations are no longer applicable in this context.

The transformed multirank \cite{ref_ttnn} of $\tX$ can be defined as a vector $\mathbf{r} \in \mathbb{R}^{n_3}$ with its $i$-th entry being the rank of $\hat{\tX}^{(i)}_\mathsf{T}$. Then, we propose the transformed tensor sum rank as follows.

\begin{mydef}[Transformed tensor sum rank]
  The transfo- rmed sum rank, $\operatorname{rank}_{sum}(\tX)$, is defined as the sum of the tensor multirank, i.e.,
  \begin{equation}
      \label{sum_rank}
      \operatorname{rank}_{sum}(\tX) = \sum_i^{n_3} r_i = \operatorname{rank}(\overline{\tX_\mathsf{T}}).
  \end{equation}
\end{mydef}

\begin{mydef}[UTNN]
  \label{def:ttsvd}
  The unitary transformed tensor nuclear norm (UTNN) of $\tX \in \mathbb{C}^{n_1 \times n_2 \times n_3}$ is defined as the nuclear norm of the block diagonal matrix in the transformed domain, i.e., 
  \begin{equation}
      \label{TTNN}
      \|\tX\|_{\mathsf{T}*} = \|\overline{\t{X}_\mathsf{T}}\|_* = \operatorname{Tr}(\overline{\t{S}_\mathsf{T}}).
  \end{equation}
\end{mydef}

Note that UTNN is derived from the dual norm of the transformed tensor spectral norm, and we have proved that UTNN is the convex envelope of the transformed tensor sum rank \eqref{sum_rank} on a unit ball of the transformed tensor spectral norm. See Appendix\ref{appd:ttnn} for more detail.

In addition, it is worth noting that our proposed transformed t-SVD can be extended to tensors of order greater than three via a recursive approach similar to how \cite{martin2013order} extended the t-product to tensors of order higher than three.

% insert network image
\begin{figure*}[t]
  \centering
  \includegraphics[scale=0.6]{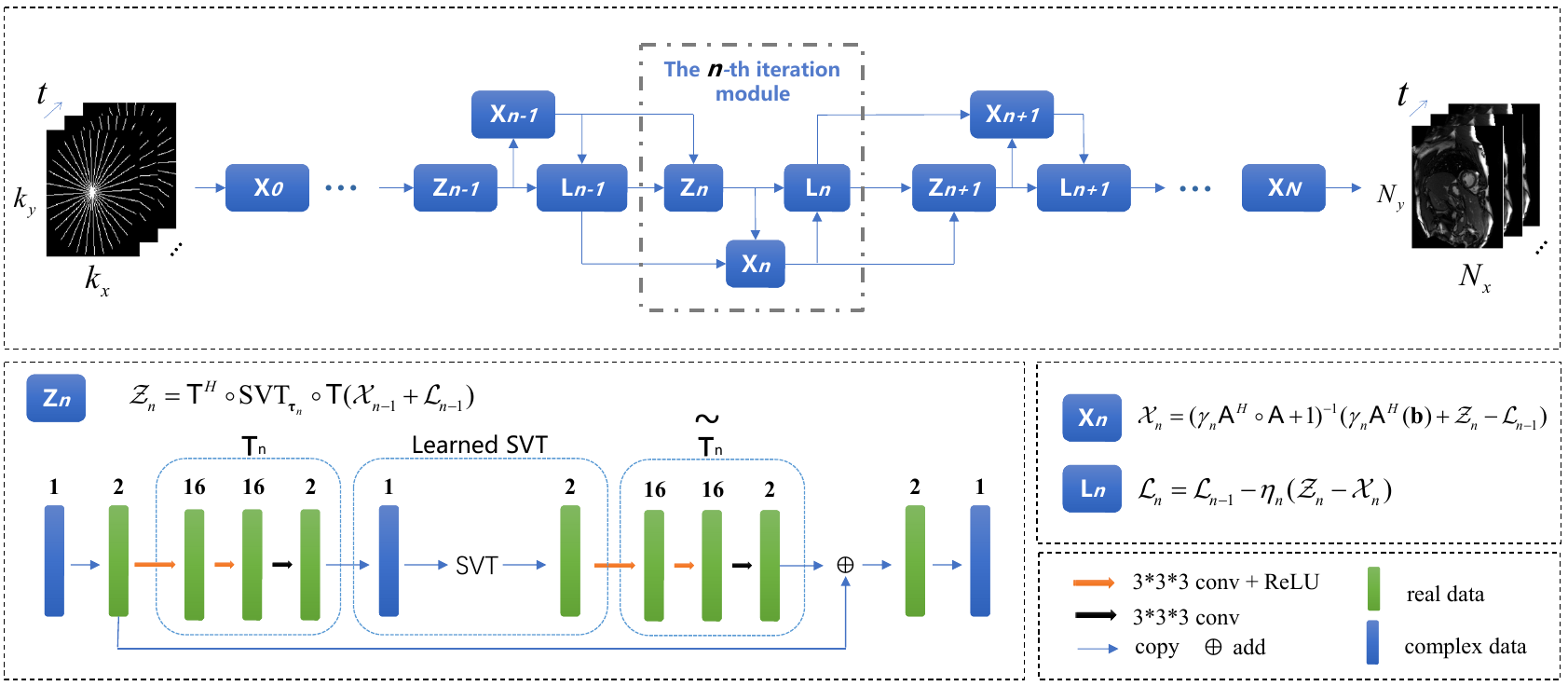}
  \caption{The proposed T$^2$LR-Net framework. The T$^2$LR-Net is an unrolling neural network that unrolls N (fixed) iteration of the algorithm \eqref{iter_alg2} into N iteration modules. Each iteration module contains three blocks: the transformed tensor low-rank prior block $\mathbf{Z}_n$, the reconstruction block $\mathbf{X}_n$, and the multiplier update block $\mathbf{L}_n$. The transformed tensor low-rank prior block is incorporated with the CNN and the hyperparameters are learned through the training process. The number above each color block represents the current channel count. The first row of the figure shows the T$^2$LR-Net framework, and the second row shows the detail of the three blocks.}\label{fig_net}
\end{figure*}

%%%%%%%%%%%%%%%%%%%%%%%%%%%%%%%%%%%%%%%%%%%%%%%%%%%%%%%%%%%%%%%%%%%%%%%%%%%%%%%%%%%%%%%%%%%%%%%%%%%%%%%%%%%%%%%%%%%%%%%%%%%%%%%%%%%%%%%%%%%%%%%%%%%%%%%%%
\subsection{UTNN-based Iterative Optimization Algorithm}
\label{subsec:ttnn_alg}

Based on UTNN of the tensor, we propose the UTNN regularized dynamic MR reconstruction model as
\begin{equation}
  \min_{\tX} \frac 12 \Vert \mathsf A(\tX)-\mathbf{b} \Vert_2^2+\lambda \Vert \tX \Vert_{\mathsf{T}*}.
\end{equation}
The above optimization problem can be converted into the following constraint problem, 
\begin{equation}
  \min_{\tX} \frac 12 \Vert \mathsf A(\tX)-\mathbf{b} \Vert_2^2+\lambda \Vert \t{Z} \Vert_{\mathsf{T}*}
  \ \ s.t. \  \t{Z} = \tX.
\end{equation}
The augmented Lagrangian function of the above optimization problem is formulated as,
\begin{equation}
  \begin{aligned}
      \label{augmentedlagrangian}
  \mathsf{L}(\tX, \t{Z}, \t{W}) = \frac 12 \Vert \mathsf A(\tX)-\mathbf{b} \Vert_2^2+\lambda \Vert \t{Z} \Vert_{\mathsf{T}*} + \\
  <\t{W}, \tX - \t{Z}> + \frac {\mu}2 \Vert \t{Z} - \tX \Vert_F^2,
  \end{aligned}
\end{equation}
where $\t{W}$ is the Lagrangian multiplier and $\mu > 0$ is the penalty parameter. After a straightforward complete-the-squares procedure for the last two terms of \eqref{augmentedlagrangian}, we have
\begin{equation}
  \begin{aligned}
      \label{augmentedlagrangian2}
  \mathsf{L}(\tX, \t{Z}, \t{L}) = \frac 12 \Vert \mathsf A(\tX)-\mathbf{b} \Vert_2^2+\lambda \Vert \t{Z} \Vert_{\mathsf{T}*} + \\
  \frac {\mu}2 \Vert \t{Z} - \tX - \t{L} \Vert_F^2 - \frac {\mu}2 \Vert \t{L} \Vert_F^2,
  \end{aligned}
\end{equation}
where $\t{L} = \frac {\t{W}} \mu$. The above can be efficiently solved with ADMM \cite{afonso2010admm}, which yields in solving the following subproblems:
\begin{align}
  & \t{Z}_{n} = \min_{\t{Z}} \lambda \Vert \t{Z} \Vert_{\mathsf{T}*} + \frac {\mu}2 \Vert \t{Z} - \tX_{n-1} - \t{L}_{n-1}\Vert_F^2, \label{z}\\
  & \tX_{n} = \min_{\tX} \frac 12 \Vert \mathsf A(\tX)-\mathbf{b} \Vert_2^2 + \frac {\mu}2 \Vert \t{Z}_{n} - \tX  - \t{L}_{n-1}\Vert_F^2, \label{x}\\
  & \t{L}_{n} = \t{L}_{n-1} - \eta (\t{Z}_{n}-\tX_{n}),
\end{align}
where the subscript $n$ denotes the $n$-th iteration and $\eta$ denotes the update rate. 

\subsubsection{$\t{Z}$ Subproblem}
The subproblem \eqref{z} can be solved by the transformed tensor singular value thresholding ($\mathsf{T}$-TSVT) operator which is detailed in Appendix\ref{appd:ttsvt},
\begin{equation}
    \label{eq21}
    \begin{aligned}
      \t{Z}_{n} &= \mathsf{T}\mathrm{-TSVT}_{{\lambda}/{\mu}}(\tX_{n-1} + \t{L}_{n-1}) \\
                &= \mathsf{T}^H \circ \operatorname{SVT}_{{\lambda}/{\mu}} \circ \mathsf{T}(\tX_{n-1} + \t{L}_{n-1}).
    \end{aligned}
\end{equation}
It moves back and forth between the original and transformed domain. The SVT denotes the matrix singular value thresholding \cite{cai2010singular} on each frontal slice of the input tensor ($[\tX_{n-1} + \t{L}_{n-1}]^{(i)}, i=1...n_t$) with the threshold ${{\lambda}/{\mu}}$.

\subsubsection{$\tX$ Subproblem}
the subproblem \eqref{x} is a quadratic problem that can be solved analytically, i.e.,
\begin{equation}
  \label{eq22}
  \tX_{n} = (\mathsf{A}^H \circ \mathsf{A} + \mu)^{-1}(\mathsf{A}^H(\mathbf{b})+\mu \t{Z}_{n} - \mu \t{L}_{n-1}).
\end{equation}
If the samples are on the Cartesian grid in the single-coil scenario, where $\mathsf A = \mathsf S \circ \mathsf{F}$ as shown in \eqref{dmri_acquisition}, the solution can be simplified by
\begin{equation}
  \label{x_cartesian}
  \tX_{n} = \mathsf{F}^H \left(\frac {\mathsf{S}^H(\mathbf{b})+\mu \mathsf{F}(\t{Z}_{n} - \t{L}_{n-1})} {\mathsf{S}^H \circ \mathsf{S} (\mathbf{1}) + \mu\ \mathbf{1}}\right),
\end{equation}
where $\mathbf{1}$ denotes an all-one tensor and the division is an element-wise operation. Note that $\mathsf{S}^H: \mathbb{C}^m \rightarrow \mathbb{C}^{n_{1} \times n_{2} \times n_{3}}$ denotes the Hermitian transpose of $\mathsf S$, and $\mathsf{S}^H(\mathbf{x})$ puts the element of the sampled vector $\mathbf{x}$ into the sampling location of the Cartesian grid and fills the rest with zeros. Meanwhile, if the acquisition is non-Cartesian and in the multi-coil scenario, the solution can be replaced by a gradient descent step \cite{huang2021deep,ref_slrnet}, i.e.,
\begin{equation}
  % \footnotesize
  \label{x_noncartesian}
  \tX_{n} = \tX_{n-1} - \beta \left[\mathsf{A}^H\circ\mathsf{A}(\tX_{n-1}) - \mathsf{A}^H(\mathbf{b}) + \mu \t{Z}_{n} - \mu \t{L}_{n-1}\right].
\end{equation}

Finally, we obtain the following iterative procedures:
\begin{equation}
  % \small
  \label{iter_alg}
  \begin{cases}
        &\t{Z}_{n} = \mathsf{T}^H \circ \operatorname{SVT}_{{\lambda}/{\mu}} \circ \mathsf{T}(\tX_{n-1} + \t{L}_{n-1}) \\
        &\tX_{n} = (\mathsf{A}^H \circ \mathsf{A} + \mu)^{-1}(\mathsf{A}^H(\mathbf{b})+\mu \t{Z}_{n} - \mu \t{L}_{n-1})\\
        &\t{L}_{n} = \t{L}_{n-1} - \eta (\t{Z}_{n}-\tX_{n})
  \end{cases}.
\end{equation}%

\subsection{The Proposed Unrolling Network: T$^2$LR-Net}
\label{subsec:net}

In the traditional optimization-based methods, the reconstruction result is obtained by iteratively solving \eqref{iter_alg}, and the hyperparameters $\lambda, \mu, \eta$ need to be selected empirically, which is usually time-consuming and un-robust. In addition, it is challenging to choose a suitable predefined unitary transform $\mathsf{T}$ in a simple formation. 

To address the aforementioned issues, we adopt the deep unrolling strategy and introduce the T$^2$LR-Net. First, we generalize the iterative solution \eqref{iter_alg} into the following scheme:
\begin{equation}
  % \small
  \label{iter_alg2}
  \begin{cases}
      \mathbf{Z}_{n}: &\t{Z}_{n} = \mathsf{T}^H \circ \operatorname{SVT}_{\bm{\tau}_{n}} \circ \mathsf{T}(\tX_{n-1} + \t{L}_{n-1}) \\
      \mathbf{X}_{n}: &\tX_{n} = (\gamma_n \mathsf{A}^H \circ \mathsf{A} + 1)^{-1}(\gamma_n \mathsf{A}^H(\mathbf{b})+\t{Z}_{n} - \t{L}_{n-1})\\
      \mathbf{L}_{n}: &\t{L}_{n} = \t{L}_{n-1} - \eta_n (\t{Z}_{n}-\tX_{n})
  \end{cases}.
\end{equation}%
The $\bm{\tau} \geq 0 \in \mathbb{R}_+^{n_t}$ replaces the single real number ${\lambda}/{\mu} \in \mathbb{R}_+$ with a tensor singular value threshold vector. The $i$-th element $\tau_{i}$ thresholds the $i$-th frontal slice of the tensor, making the thresholding more flexible. We replace $1/\mu$ with $\gamma$ to avoid numerical instability. This is because, in the Cartesian sampling cases \eqref{x_cartesian}, if the parameter $\mu$ is learned to be zero, the unsampled $k$-space position would result in $\frac{0}{0}$, which is numerically unstable.
The hyperparameters are denoted by a subscript $n$ (e.g., $\bm{\tau}_{n}$), indicating that these hyperparameters vary from iteration to iteration. Note that the unitary transform $\mathsf{T}_n$ differs across iterations to leverage different transformed tensor low-rank priors to enhance reconstruction performance.

The above iterative procedure \eqref{iter_alg2} is then unrolled into the proposed T$^2$LR-Net. The pseudocode of the T$^2$LR-Net is shown in Alg.\ref{algo_tlr}.
The framework of the network is shown in Figure \ref{fig_net}. T$^2$LR-Net unrolls the iterative steps of the UTNN-based optimization algorithm \eqref{iter_alg2} into $N$ iteration modules. Each module contains three blocks corresponding to the three subproblems in \eqref{iter_alg2}, i.e., the transformed tensor low-rank prior block $\mathbf{Z}_n$, the reconstruction block $\mathbf{X}_n$, and the multiplier update block $\mathbf{L}_n$. 
The CNN is utilized in the $\mathbf{Z}_n$ block to adaptively learn the transformation from the dynamic MR image datasets, and the hyperparameters ($\tau, \gamma, \eta$) are also automatically learned through the training process. 
The three blocks are described in detail as follows:

\IncMargin{1em}
\begin{algorithm} 
	\SetKwData{Left}{left}
	\SetKwData{This}{this}
	\SetKwData{Up}{up} 
	\SetKwFunction{Union}{Union}
	\SetKwFunction{FindCompress}{FindCompress} 
	\SetKwInOut{Require}{Require}
	\SetKwInOut{Initialize}{Initialize}
	\SetKwInOut{Return}{Return}
	
	\Require{$\{\mathsf{T}_n,\widetilde{\mathsf{T}_n},\mathbf{\tau}_n,\gamma_n,\eta_n: n=1,\dots,N\}$} 
	\Initialize{$\mathcal{X}_0 = \mathsf{A}^H\mathbf{b}$, $\mathcal{Z}_0 = \mathbf{0}$, $\mathcal{L}_0 = \mathbf{0}$}
	 \BlankLine 
	 
	%  \emph{special treatment of the first line}\; 
	 \For{$n=1,\dots,N$}{ 
		\emph{Transformed Tensor Low-rank Prior Block}:\\
		$\t{Z}_{n} = \mathsf{T}^H \circ \operatorname{SVT}_{\bm{\tau}_{n}} \circ \mathsf{T}(\tX_{n-1} + \t{L}_{n-1})$ \;
    \quad \\
		\emph{Reconstruction Block}:\\
		$\tX_{n} = (\gamma_n \mathsf{A}^H \circ \mathsf{A} + 1)^{-1}(\gamma_n \mathsf{A}^H(\mathbf{b})+\t{Z}_{n} - \t{L}_{n-1})$\;
    \quad \\
		\emph{Multiplier Update Block}:\\
		$\t{L}_{n} = \t{L}_{n-1} - \eta_n (\t{Z}_{n}-\tX_{n})$
 	 } 
	  \BlankLine 
	\Return{$\mathcal{X}_{N}$}
      \caption{T$^2$LR-Net}
 	 	  \label{algo_tlr} 
 	 \end{algorithm}
 \DecMargin{1em} 

\subsubsection{Transformed Tensor Low-rank Prior Block $\mathbf{Z}_n$}
\label{subsubsec:z}
In this block, a learned $\mathsf{T}$-TSVT related to the $\t{Z}_{n}$ subproblem in \eqref{iter_alg2} \eqref{eq21} is embedded into the neural network,
\begin{equation}
  \label{eq_zblock}
  \begin{aligned}
      \t{Z}_{n} &= \mathsf{T}_n^H \circ \operatorname{SVT}_{\bm{\tau}_{n}}\circ\mathsf{T}_n(\tX_{n-1} + \t{L}_{n-1}) \\
                &\approx \widetilde{\mathsf{T}_n} \circ \operatorname{SVT}_{\bm{\tau}_{n}}\circ\mathsf{T}_n(\tX_{n-1} + \t{L}_{n-1}).
  \end{aligned}
\end{equation}
Specifically, the transform $\mathsf{T}_n$ and the Hermitian transpose transform $\mathsf{T}_n^H$ are learned by two different CNNs. Instead of the traditional SVT, a learned matrix SVT operator \cite{ref_slrnet} is incorporated to adaptively determine the threshold ${\tau}_{n,i}$ of every frontal slice of the transformed image, i.e., the threshold vector $\bm{\tau}_{n}$.
For a certain frontal slice $\hat{\t{Y}}_\mathsf{T}^{(i)} = \mathsf{T}_n(\tX_{n-1} + \t{L}_{n-1})=\mathbf{U}\mathbf{S}\mathbf{V}^H$,
the learned matrix SVT is based on the following scheme:
\begin{equation}
  \begin{cases}
      \operatorname{SVT}_{\tau_{n,i}}(\hat{\t{Y}}_\mathsf{T}^{(i)})&= \mathbf{U} \mathbf{S}_{\tau_{n,i}} \mathbf{V}^H \\
      \mathbf{S}_{\tau_{n,i}}&= \operatorname{ReLU}(\mathbf{S} - \tau_{n,i}) \\
      \tau_{n,i}&= \operatorname{sigmoid}(a_{n,i}) \cdot \max(\mathbf{S})
  \end{cases}.
\end{equation}
In the above equation, $\operatorname{ReLU}$ denotes the rectifier linear units \cite{glorot2011relu}, $\operatorname{sigmoid}$ denotes the sigmoid activate function, and $a_{n,i}$ is set as a learnable network parameter with an initial value of -2. 

Due to the complexity of determining the Hermitian transpose of a CNN-learned transform, we avoid strictly using the Hermitian transpose transform $\mathsf{T}_n^H$. Instead, we build another CNN with the same structure, denoted as $\widetilde{\mathsf{T}_n}$, to approximate $\mathsf{T}_n^H$.
In addition, instead of imposing the unitary or inverse constraints on the two CNNs in the loss function \cite{ref_slrnet,ref_ISTANET}, we allow them to adaptively learn priors from dynamic MR datasets. In this way, the network can utilize both explicit low-rank properties and implicit deep image priors to enhance reconstruction performance. We will provide a more detailed discussion in Section \ref{subsec:u_lr}.

\subsubsection{Reconstruction Block $\mathbf{X}_n$}
This block corresponds to the $\t{X}_{n}$ subproblem in \eqref{iter_alg2} and generates the reconstruction results. We set $\gamma_n = \operatorname{ReLU}(\hat{\gamma}_n)$, where $\hat{\gamma}_n$ is learnable with an initial value of 0.1. 

\subsubsection{Multiplier Update Block $\mathbf{L}_n$}
This block corresponds to the $\t{L}_{n}$ subproblem in \eqref{iter_alg2} and is used to update the Lagrange multiplier. We define $\eta_n = \operatorname{ReLU}(\hat{\eta}_n)$, where $\hat{\eta}_n$ is learnable with an initial value of 1.

% \vspace{-10pt}
\subsection{Loss Function}
We use the mean squared error (MSE) as the loss function of T$^2$LR-Net, which is defined as
{\small
\begin{equation}
  \label{loss_func}
      Loss = \sum_{(\bar{\tX}, \mathbf{b}) \in \Omega}\Vert \bar{\tX} - f_{net}(\mathbf{b}|\theta) \Vert_F^2.
\end{equation}}%
In the above equation, $\Omega$ denotes the given training data, $\bar{\tX}$ is a fully sampled ground-truth data, $\mathbf{b}$ is the undersampled $k$-space data, $f_{net}$ denotes the output of the network, and $\theta$ is the learnable parameters of the network, which include $\bm{\tau}_n$, $\gamma_n$, $\eta_n$, as well as the CNN-learned transformations $\mathsf{T}_n$ and $\widetilde{\mathsf{T}_n}$.

% \vspace{-10pt}
\subsection{Implementation Details} 
\label{subsec:implementation}

Taking into account the trade-off between computational burden and reconstruction performance, we choose 15 iteration modules to compose the proposed T$^2$LR-Net. To facilitate training of the embedded CNN in $\mathbf{Z}_n$ blocks, we split each input complex-valued data into two real-valued channels. The CNNs for learning $\mathsf{T}_n$ and $\widetilde{\mathsf{T}_n}$ consist of three convolutional layers. The first two layers have 16 channels followed by a ReLU operator. The third layer comprises 2 channels and does not use a ReLU activation function to avoid truncating the negative part of the output. The convolution kernels of all the convolutional layers are of the size $3\times3\times3$, and the stride is 1. He initialization \cite{he2015initial} is used for the convolutional layers.

During training, we adopt a pseudo-random mask strategy instead of using a fixed sampling mask throughout the training process, as seen in previous training schemes \cite{ref_DCCNN,ref_slrnet}. At each training step, we generate a random sampling mask of a particular type and acceleration (e.g., a radial sampling mask with 16 randomly selected lines). This approach allows the network to learn deep features more adaptively and avoid overfitting to a fixed sampling mask. 

The model is implemented in the framework of Tensorflow \cite{abadi2016tensorflow}. The batch size is set as 1. The Adam optimizer \cite{kingma2014adam} with parameters $\beta_1=0.9,\ \beta_2=0.999$ and $\epsilon=10^{-8}$ is adopted, and the exponential decay learning rate \cite{zeiler2012adadelta} is used with an initial learning rate of 0.001 and a decay of 0.95. All the experiments are performed on a workstation with Intel Xeon W-2123 CPU and NVIDIA Tesla GV100 GPU (32 GB memory). The source code of our method is available at \url{https://github.com/yhao-z/T2LR-Net}.

\subsection{Complexity and Convergence Analysis}
\label{subsec:complexity}

Regarding the computational complexity of the model, assuming the size of the input tensor image $\mathcal{X}$ is $H \times W \times T$ with $H > W$. 
For the $Z_n$ subproblem in \eqref{iter_alg2}, the total complexity of the CNNs (the transformation $T$ and its transpose) is $HWT \times 3^3 \times 16 \times (16 + 2 + 2) \times 2$, i.e., $\mathcal{O}(HWT)$. Here, we substitute the kernel size and the number of channels from Subsec.\ref{subsec:implementation} to simplify the calculation. The cost of the SVT operator is dominated by tensor SVD, whose complexity is $\mathcal{O}(W^2HT)$.
For the $X_n$ subproblem, in the Cartesian sampling scenario \eqref{x_cartesian}, its cost is dominated by linear operations and fast Fourier transforms, resulting in $\mathcal{O}(HW \log(HW)T)$ computational complexity.
The $L_n$ subproblem only includes linear operations, resulting in the cost of $\mathcal{O}(HWT)$.
Therefore, for $N$ iteration modules, the total computational complexity should be $\mathcal{O}(NW^2HT)$.

Regarding the convergence of \eqref{iter_alg2}, since this algorithm shares the same form as the fundamental 2-block ADMM, its convergence can be guaranteed by Section 3.1 of reference \cite{boyd2011distributed}. Furthermore, in the case of single-coil Cartesian sampling, as each substep of ADMM has an exact analytical solution, assurance of the convergence rate of this algorithm to $\mathcal{O}(1/K)$ can be obtained from references \cite{he20121, monteiro2013iteration}, where K denotes the number of iterations.

\section{Experiments and Results}
\label{sec:experiments}

\subsection{Dataset}
% \subsubsection{Dataset}
\label{subsubsec:dataset}

We evaluate the T$^2$LR-Net using two cardiac cine MRI datasets. The first one is the open-access OCMR dataset \cite{ref_ocmr}, while the other, named TCMR, is established based on the work of Tsotsos \emph{et al.} \cite{andreopoulos2008efficient} and previously evaluated in \cite{zheng2019cascaded}.

The OCMR dataset contains 78 fully sampled raw data collected on 3T Siemens MAGNETOM Prisma machine and 126 fully sampled raw data on 1.5T Siemens Avanto and Sola machine. We choose 68 of the 3T data for training, and select the rest 10 of the 3T data and the other 12 data from the 1.5T Siemens Sola machine for testing to evaluate the robustness and generalization ability of the proposed network. We crop the training data into the size of $144 \times 112 \times 16$ ($x \times y \times t$), and the strides along three dimensions are 15, 15 and 7, respectively. Finally, we obtain 1099 training data and 22 uncropped test data. The coil sensitivity maps are computed by ESPIRiT \cite{ref_multicoil}. Both single-coil and multi-coil data were used for the experiments. The raw multi-coil data of each frame were combined using the coil sensitivity maps estimated by ESPIRiT to produce single-coil complex-valued data.

The TCMR dataset is comprised of short-axis cardiac cine MR images from 33 subjects. Although the images may not match a realistic experiment as the raw data, they are still useful as an auxiliary dataset to evaluate the efficacy of methods. The images are scanned with a GE Genesis Signa MR scanner using the FIESTA scan protocol. A total of 399 slices with each slice of 256$\times$256$\times$20 ($x \times y \times t$) are collected. We crop the images into the size of $128 \times 128 \times 16$ with the strides of 15, 15, and 7 along $x$, $y$, and $t$ dimensions, respectively. Finally, we obtain an augmented dataset consisting of 3782 cardiac cine MR images. We use 3392 images from 30 subjects for training and 390 images from the rest 3 subjects for testing.

\subsection{Experimental settings}
To demonstrate the efficacy of the T$^2$LR-Net in dynamic MR cine imaging, we conduct comprehensive experiments involving single-coil/multi-coil scenarios, retrospective/prospective reconstruction, different acceleration factors, and different sampling patterns, as shown in Table \ref{tab:secsetting}. In retrospective experiments, two commonly used sampling patterns are considered, i.e., the Cartesian pseudo-radial sampling pattern \cite{ref_ktslr} and the variable density random sampling (Vds) pattern. Three different sampling cases are considered for each sampling pattern, where the radial sampling pattern involves 8, 16, and 30 lines, and the Vds pattern involves 8, 10, and 12 acceleration rates (acc = the total number of pixels / the number of the sampled pixels). In prospective experiments, we evaluate the proposed T$^2$LR-Net using the VISTA sampling pattern \cite{ahmad2015vista} since the undersampled raw data from OCMR are acquired by VISTA.

\begin{figure}[htbp]
  \centering
  \includegraphics[scale=0.50]{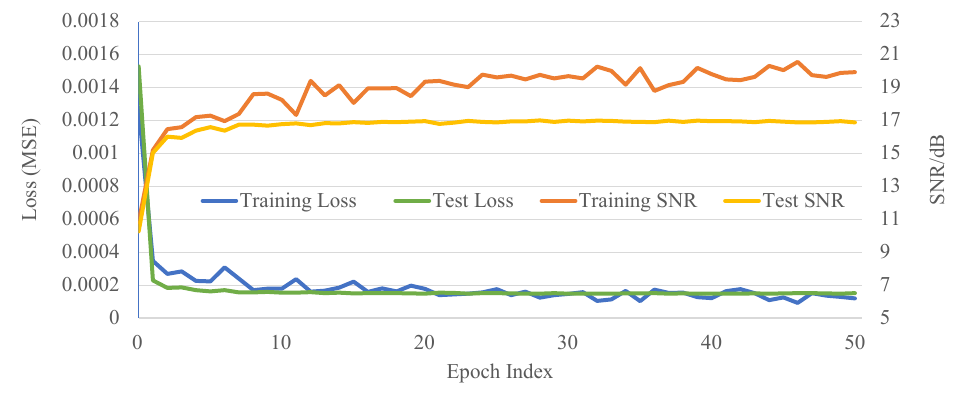}
  \caption{{The training and test loss curves of the proposed T$^2$LR-Net.}}
  \label{fig_loss}
\end{figure}

We use signal-to-noise ratio (SNR) and structural similarity (SSIM) \cite{wang2004image} to evaluate the results. In each scenario, we have carefully tuned the parameters of all the compared optimization-based methods to ensure optimal performance. The unrolling network-based methods are retrained properly using the corresponding dataset for 50 epochs to guarantee fair comparisons. Figure \ref{fig_loss} shows the training and test loss curves, as well as the SNR curves, for the proposed T$^2$LR-Net under the 10-fold Vds pattern on the OCMR dataset. The training and test loss curves under other sampling patterns are similar and are omitted here for brevity. It can be observed that the training has converged after 50 epochs, and there is no degradation in the reconstructed SNR for the test set, indicating that there is no overfitting.

% Table generated by Excel2LaTeX from sheet 'Sheet4'
\begin{table}[htbp]
  \centering
  
  \caption{Experimental settings of each section.}
    \resizebox{1\linewidth}{!}{\begin{tabular}{cccc}
    \toprule
    \textbf{Section} & \textbf{\#Coils} & \textbf{Retro./Prosp.} & \textbf{Sampling-Acceleration/Lines} \\
    \midrule
    4.3  & Single & Retrospective & Radial-8,16,30 , Vds-8,10,12 \\
    4.4  & Multi & Retrospective & Radial-16, Vds-10 \\
    4.5  & Multi & Prospective & VISTA-8 \\
    5.1-4 & Single & Retrospective & Radial-16 \\
    \bottomrule
    \end{tabular}}%
  \label{tab:secsetting}%
  
\end{table}%
% \vspace{-5pt}
\subsection{Retrospective Experiments in single-coil scenario}
We compare the reconstruction results of the proposed T$^2$LR-Net with four state-of-the-art (SOTA) optimization-based methods (TNN \cite{ref_tnn1}, F2TNN\cite{ref_pre_ttnn}, k-t SLR\cite{ref_ktslr} and SNNTV\cite{ref_wmnn2}) and two unrolling network-based methods (DCCNN \cite{ref_DCCNN}, SLR-Net\cite{ref_slrnet}). Note that TNN and F2TNN are based on the t-SVD framework with predefined transforms, and these two methods are originally applied in tensor completion tasks. We adapt them to the dynamic MR image reconstruction by ourselves.

\begin{figure*}[h!]
  % \vspace*{-10pt}
  \centering
  \includegraphics[scale=0.6]{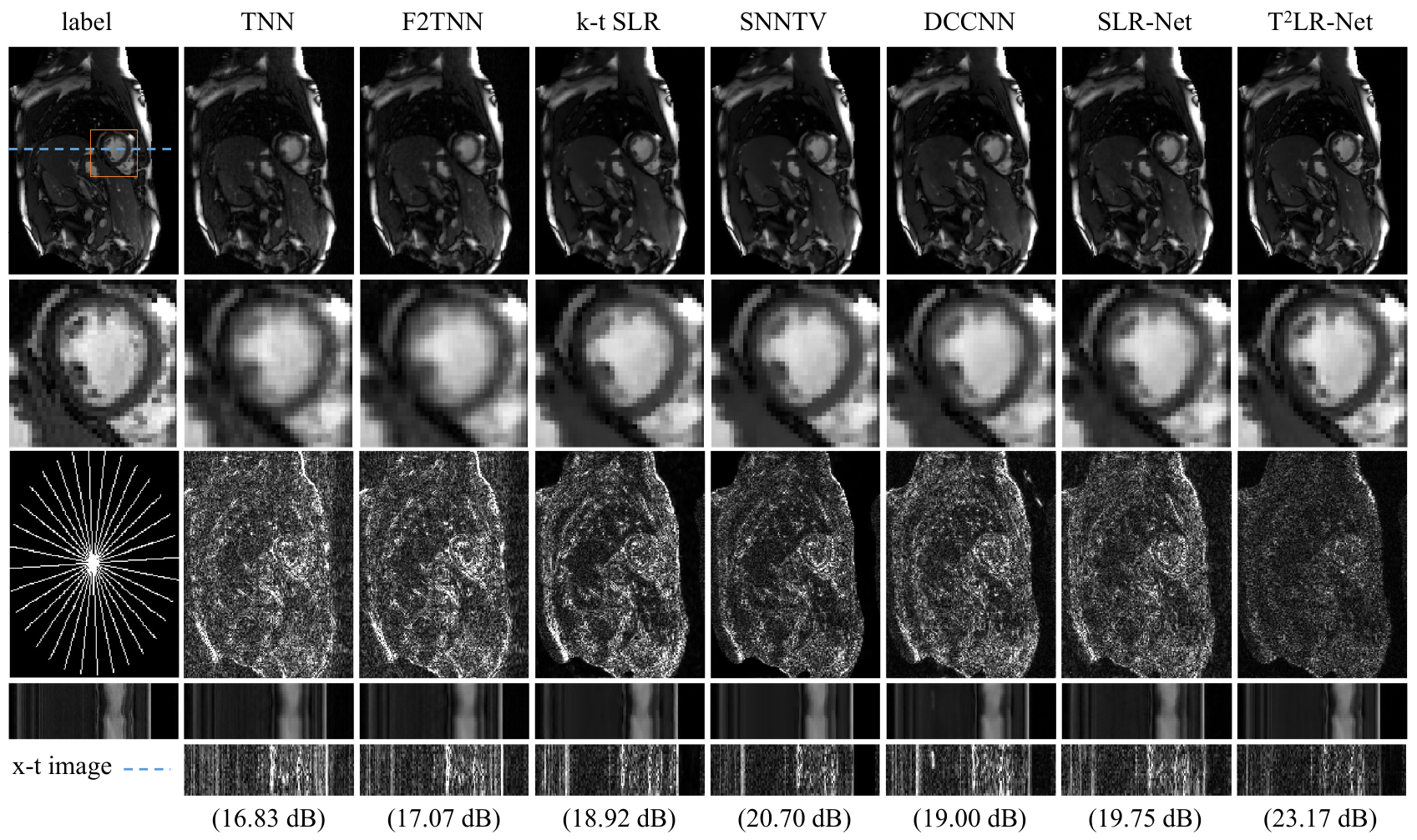}
  \caption{The OCMR reconstruction results of different methods under the pseudo-radial sampling pattern \cite{ref_ktslr} with 16 lines in the single-coil scenario. The first row shows the reconstruction images of the different methods, and the second row shows the enlarged view of the heart regions marked by the orange box. The first image in the third row displays the sampling mask, while the other images show the reconstruction error maps w.r.t. the different methods. The fourth row and the fifth row show the x-t images indicated by the blue dot line and their reconstruction error maps. The reconstruction SNRs of different methods are listed in parentheses. }
  \label{result1}
  \centering
  \includegraphics[scale=0.6]{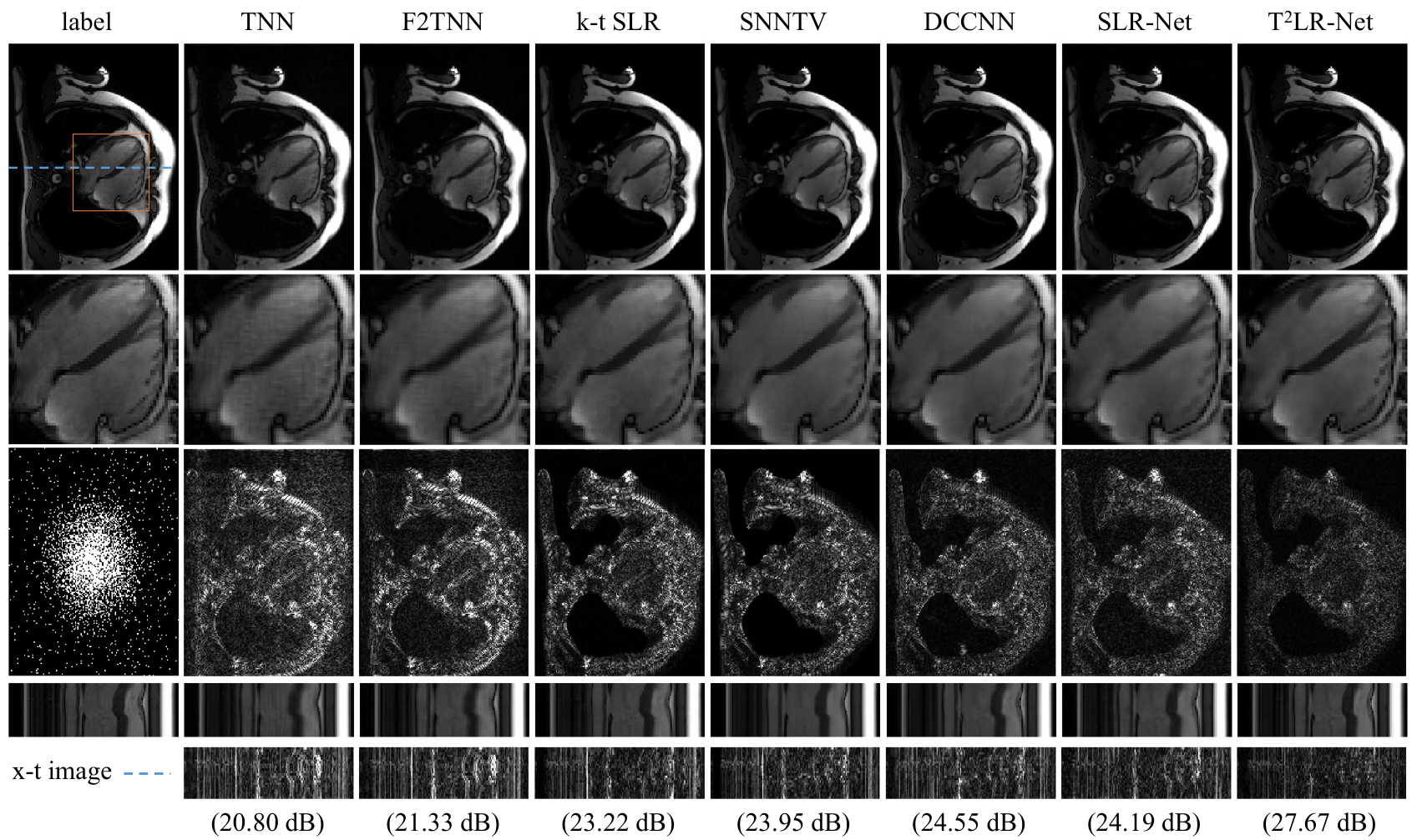}
  \caption{The OCMR reconstruction results of different methods under the variable density random sampling pattern with the 8-fold acceleration in the single-coil scenario. }
  \label{result2}
\end{figure*}

Figure \ref{result1} presents the reconstruction results for a test image from the OCMR dataset using the pseudo-radial sampling pattern \cite{ref_ktslr} with 16 lines, and Figure \ref{result2} shows the results of OCMR under the variable density random sampling pattern with the 8-fold acceleration. We report the quantitative metrics of different methods under six different sampling cases on the OCMR and TCMR datasets in Table \ref{tab_ocmr} and Table \ref{tab_tcmr}, respectively. The visualization results demonstrate that our network can provide clearer edges, finer textures, and lower errors. Quantitative analysis results indicate that the T$^2$LR-Net achieves the highest SNR and SSIM metrics. Moreover, compared to SOTA unrolling networks, it possesses the fewest parameters and comparable training and inference time.

The TNN and F2TNN methods utilize the tensor low-rank priors in predefined transformed domains. However, artificially designed transformations may not be suitable for dynamic MR images, lacking flexibility and consequently leading to suboptimal reconstruction results. Both k-t SLR and SNNTV methods are composite approaches that combine low-rank characteristics and total variation priors, achieving better results than sole low-rank methods. However, their computational complexity significantly increases, and the required convergence time significantly increases. This poses challenges for practical clinical applications. DCCNN and SLR-Net, as SOTA deep unfolding structures, have demonstrated satisfying reconstruction results. However, DCCNN relies on CNNs to extract implicit image priors, lacking physical motivation and thus facing challenges in achieving efficient and accurate training. While SLR-Net combines low-rank and sparse composite priors and utilizes them in the deep unrolling framework, its utilization of matrix low-rank prior disrupts the inherent tensor structure of dynamic MR images. Our proposed network, starting from tensor theory and leveraging CNNs to extract the tensor low-rank features, fully exploits the inherent high-dimensional structure of dynamic MR images, resulting in optimal reconstruction performance.

\begin{table*}[h]
  
  \centering
  \caption{The quantitative metrics of different methods under two sampling patterns and six different sampling cases on the test dataset of the OCMR dataset (Mean ± Standard deviation).}
  \resizebox{1\linewidth}{!}{\begin{tabular}{ccc|ccccccc}
    \toprule
          &       &       & \textbf{TNN} & \textbf{F2TNN} & \textbf{k-t SLR} & \textbf{SNNTV} & \textbf{DCCNN} & \textbf{SLR-Net} & \textbf{T$^2$LR-Net} \\
    \midrule
    \multirow{6}[6]{*}{\textbf{Radial / lines}} & \multirow{2}[2]{*}{\textbf{8}} & SNR   & 12.39 ± 2.36 & 12.43 ± 2.25 & 15.37 ± 2.14 & 17.05 ± 2.28 & 15.95 ± 1.79 & 16.14 ± 1.87 & \textbf{19.12 ± 2.43} \\
          &       & SSIM  & 0.752 ± 0.060 & 0.778 ± 0.052 & 0.910 ± 0.032 & 0.927 ± 0.028 & 0.903 ± 0.027 & 0.892 ± 0.025 & \textbf{0.944 ± 0.026} \\
\cmidrule{2-10}          & \multirow{2}[2]{*}{\textbf{16}} & SNR   & 16.35 ± 2.68 & 16.57 ± 2.53 & 18.54 ± 1.91 & 19.93 ± 2.32 & 18.53 ± 1.64 & 19.24 ± 1.76 & \textbf{22.44 ± 2.60} \\
          &       & SSIM  & 0.859 ± 0.053 & 0.882 ± 0.044 & 0.947 ± 0.021 & 0.952 ± 0.023 & 0.934 ± 0.022 & 0.939 ± 0.019 & \textbf{0.968 ± 0.020} \\
\cmidrule{2-10}          & \multirow{2}[2]{*}{\textbf{30}} & SNR   & 19.66 ± 3.07 & 19.83 ± 2.92 & 21.07 ± 2.31 & 22.10 ± 2.50 & 21.44 ± 2.25 & 21.27 ± 2.29 & \textbf{24.08 ± 2.92} \\
          &       & SSIM  & 0.923 ± 0.041 & 0.937 ± 0.035 & 0.966 ± 0.018 & 0.964 ± 0.021 & 0.961 ± 0.021 & 0.953 ± 0.019 & \textbf{0.974 ± 0.019} \\
    \midrule
    \multirow{6}[6]{*}{\textbf{Vds / acc}} & \multirow{2}[2]{*}{\textbf{8}} & SNR   & 17.24 ± 2.32 & 17.66 ± 2.35 & 18.83 ± 2.09 & 19.94 ± 2.21 & 20.37 ± 1.86 & 20.12 ± 1.73 & \textbf{22.89 ± 2.63} \\
          &       & SSIM  & 0.893 ± 0.038 & 0.923 ± 0.031 & 0.954 ± 0.020 & 0.953 ± 0.023 & 0.955 ± 0.020 & 0.953 ± 0.017 & \textbf{0.971 ± 0.018} \\
\cmidrule{2-10}          & \multirow{2}[2]{*}{\textbf{10}} & SNR   & 15.39 ± 1.89 & 15.45 ± 1.90 & 16.23 ± 1.95 & 16.50 ± 1.94 & 16.14 ± 1.80 & 16.05 ± 1.64 & \textbf{16.89 ± 1.69} \\
          &       & SSIM  & 0.901 ± 0.032 & 0.912 ± 0.030 & 0.938 ± 0.024 & 0.934 ± 0.027 & 0.930 ± 0.025 & 0.927 ± 0.024 & \textbf{0.942 ± 0.024} \\
\cmidrule{2-10}          & \multirow{2}[2]{*}{\textbf{12}} & SNR   & 13.84 ± 1.74 & 13.89 ± 1.75 & 14.65 ± 1.88 & 14.91 ± 1.83 & 14.48 ± 1.64 & 14.40 ± 1.52 & \textbf{15.12 ± 1.66} \\
          &       & SSIM  & 0.877 ± 0.035 & 0.887 ± 0.033 & 0.921 ± 0.027 & 0.918 ± 0.030 & 0.910 ± 0.028 & 0.906 ± 0.028 & \textbf{0.924 ± 0.027} \\
    \midrule
    \multicolumn{3}{c|}{\textbf{Parameter}} & —     & —     & —     & —     & 954340 & 293808 & \textbf{259244} \\
    \midrule
    \multicolumn{3}{c|}{\textbf{Approximate training time/h}} & —     & —     & —     & —     & \textbf{9.2} & 16.7  & 22.9 \\
    \midrule
    \multicolumn{3}{c|}{\textbf{Inference time/s}} & 20.5  & 215.6 & 364.8 & 3472.8 & \textbf{0.5} & 0.8   & 1.3 \\
    \bottomrule
    \end{tabular}}%
    \label{tab_ocmr}%
\end{table*}%

\begin{table}[h]
  
  \centering
  \caption{The quantitative metrics of different methods under two sampling patterns and six different sampling cases on the test dataset of the TCMR dataset (Mean ± Standard deviation).}
  \resizebox{1\linewidth}{!}{\begin{tabular}{ccc|ccc}
    \toprule
          &       &       & \textbf{DCCNN} & \textbf{SLR-Net} & \textbf{T$^2$LR-Net} \\
    \midrule
    \multirow{6}[6]{*}{\textbf{Radial / lines}} & \multirow{2}[2]{*}{\textbf{8}} & SNR   & 25.19±3.08 & 23.71±2.89 & \textbf{25.91±3.45} \\
          &       & SSIM  & 0.948±0.027 & 0.936±0.031 & \textbf{0.953±0.027} \\
\cmidrule{2-6}          & \multirow{2}[2]{*}{\textbf{16}} & SNR   & 28.89±3.39 & 26.72±2.99 & \textbf{29.29±3.61} \\
          &       & SSIM  & 0.973±0.017 & 0.960±0.020 & \textbf{0.974±0.018} \\
\cmidrule{2-6}          & \multirow{2}[2]{*}{\textbf{30}} & SNR   & 32.46±3.43 & 30.36±3.16 & \textbf{33.27±3.73} \\
          &       & SSIM  & 0.987±0.010 & 0.981±0.012 & \textbf{0.988±0.010} \\
    \midrule
    \multirow{6}[6]{*}{\textbf{Vds / acc}} & \multirow{2}[2]{*}{\textbf{8}} & SNR   & 29.36±3.00 & 28.83±2.84 & \textbf{29.58±3.13} \\
          &       & SSIM  & 0.976±0.015 & 0.974±0.016 & \textbf{0.977±0.015} \\
\cmidrule{2-6}          & \multirow{2}[2]{*}{\textbf{10}} & SNR   & 26.85±2.48 & 25.78±2.34 & \textbf{27.06±2.50} \\
          &       & SSIM  & 0.967±0.018 & 0.961±0.019 & \textbf{0.969±0.017} \\
\cmidrule{2-6}          & \multirow{2}[2]{*}{\textbf{12}} & SNR   & 24.63±2.23 & 24.30±2.18 & \textbf{24.78±2.26} \\
          &       & SSIM  & 0.952±0.022 & 0.950±0.023 & \textbf{0.953±0.022} \\
    \bottomrule
    \end{tabular}}%
    \label{tab_tcmr}%
\end{table}%

Additionally, we compared three SOTA transform-based TNN methods: FTNN \cite{jiang2020framelet}, S2NTNN \cite{luo2022self}, and TNN-data \cite{ref_ttnn}. These methods employ one-dimensional transformations on the temporal dimension: FTNN uses a framelet transformation, S2NTNN employs a fully connected network to learn the transformation in an unsupervised manner, and TNN-data first obtains the reconstruction result using the FFT-TNN method, then utilizes the SVD of the reconstructed image to obtain a data-driven transformation. Simultaneously, we compared the results between the same transformation (T$^2$LR-Net-shared, achieved by sharing the parameters of CNNs between different iteration modules) and different transformations (proposed) within multiple iteration modules in T$^2$LR-Net. The reconstruction results for the five mentioned methods under radial-16 sampling on the OCMR test data (same as in Fig.\ref{result1}) are presented in Tab.\ref{tab:tnn_cmp}. The table also includes the average and standard deviation of SNR and SSIM for both T$^2$LR-Net-shared and T$^2$LR-Net across the entire test dataset. The comparison reveals that our proposed CNN-learned transform-based TNN, with its advantages of multi-dimensional transformation and supervised learning, effectively adapts to the task of dynamic MR reconstruction. In contrast, other transform-domain TNN methods, while achieving good results in tensor completion, struggle to accommodate the demands of MRI reconstruction. This challenge arises due to the frequency-domain data acquisition in MRI, introducing difficulties in reconstruction across different domains, whereas tensor completion techniques often only consider the image domain. The decreased reconstruction accuracy in T$^2$LR-Net-shared suggests that utilizing different transformed domains across different iterative modules enhances the network's learning capacity and its ability to exploit the low-rank properties, ultimately improving reconstruction accuracy.

\begin{table}[htbp]
  \centering
  \caption{The quantitative results compared to SOTA transform-based TNN methods under radial-16 sampling on the OCMR test data (same as that in Fig.\ref{result1}). The table also includes the average and standard deviation of SNR and SSIM for both T$^2$LR-Net-shared and T$^2$LR-Net across the entire test dataset.}\label{tab:tnn_cmp}
  \resizebox{1\linewidth}{!}{\begin{tabular}{ccccccc}
    \toprule
          & \multicolumn{2}{c}{\textbf{FTNN}} & \multicolumn{2}{c}{\textbf{S2NTNN}} & \multicolumn{2}{c}{\textbf{TNN-data}} \\
    \midrule
    \multicolumn{1}{l}{SNR/SSIM} & \multicolumn{2}{c}{11.12/0.736} & \multicolumn{2}{c}{11.37/0.798} & \multicolumn{2}{c}{16.12/0.875} \\
    \midrule
          & \multicolumn{3}{c}{\textit{\textbf{T$^2$LR-Net-shared}}} & \multicolumn{3}{c}{\textit{\textbf{T$^2$LR-Net}}} \\
    \midrule
    \multirow{2}[2]{*}{SNR/SSIM} & \multicolumn{3}{c}{22.21/0.975} & \multicolumn{3}{c}{23.17/0.979} \\
          & \multicolumn{3}{c}{21.14±2.49/0.961±0.021} & \multicolumn{3}{c}{22.44±2.60/0.968±0.020} \\
    \bottomrule
    \end{tabular}}%
\end{table}%

During the process of signal acquisition for MRI, noise is often mixed in the data. Therefore, we compared the reconstruction results of different methods under various noise levels. We conducted experiments using the OCMR dataset with the Radial-16 sampling pattern. Gaussian white noise with signal-to-noise ratios of 30dB and 20dB was added to the acquired undersampled $k$-space data. The reconstruction results on the OCMR test data (same as that in Fig.\ref{result1}) are shown in Table \ref{tab:noise}. Additionally, we provide the average quantitative metrics and standard deviations for both SLR-Net and T$^2$LR-Net across the entire test dataset. The results indicate that our proposed T$^2$LR-Net achieves superior reconstruction performance under different noise levels compared to the SOTA methods, demonstrating its robustness to noise.

\begin{table*}[htbp]
  \centering
  \caption{The quantitative results in noisy cases under radial-16 sampling on the OCMR test data (same as in Fig.\ref{result1}). The results are shown in SNR/SSIM. The table also includes the average and standard deviation of SNR and SSIM for both SLR-Net and T$^2$LR-Net across the entire test dataset.}\label{tab:noise}
  \begin{tabular}{ccccc}
    \toprule
          & \textbf{TNN} & \textbf{k-t SLR} & \textbf{SLR-Net} & \textbf{T$^2$LR-Net} \\
    \midrule
    \multirow{2}[2]{*}{noisy-30dB} & \multirow{2}[2]{*}{15.24/0.818} & \multirow{2}[2]{*}{17.43/0.927} & 18.60/0.936 & \textbf{21.00/0.967} \\
          &       &       & 18.22±1.46/0.930±0.019 & \textbf{20.47±1.80/0.956±0.020} \\
    \midrule
    \multirow{2}[2]{*}{noisy-20dB} & \multirow{2}[2]{*}{11.07/0.654} & \multirow{2}[2]{*}{12.94/0.757} & 15.57/0.890 & \textbf{17.52/0.936} \\
          &       &       & 15.45±1.16/0.890±0.020 & \textbf{17.43±1.34/0.928±0.021} \\
    \bottomrule
    \end{tabular}%
\end{table*}%

\subsection{Retrospective Experiments in multi-coil scenario}
\label{subsec:multi_coil}
\begin{figure}[h]
  \centering
  \includegraphics[scale=0.42]{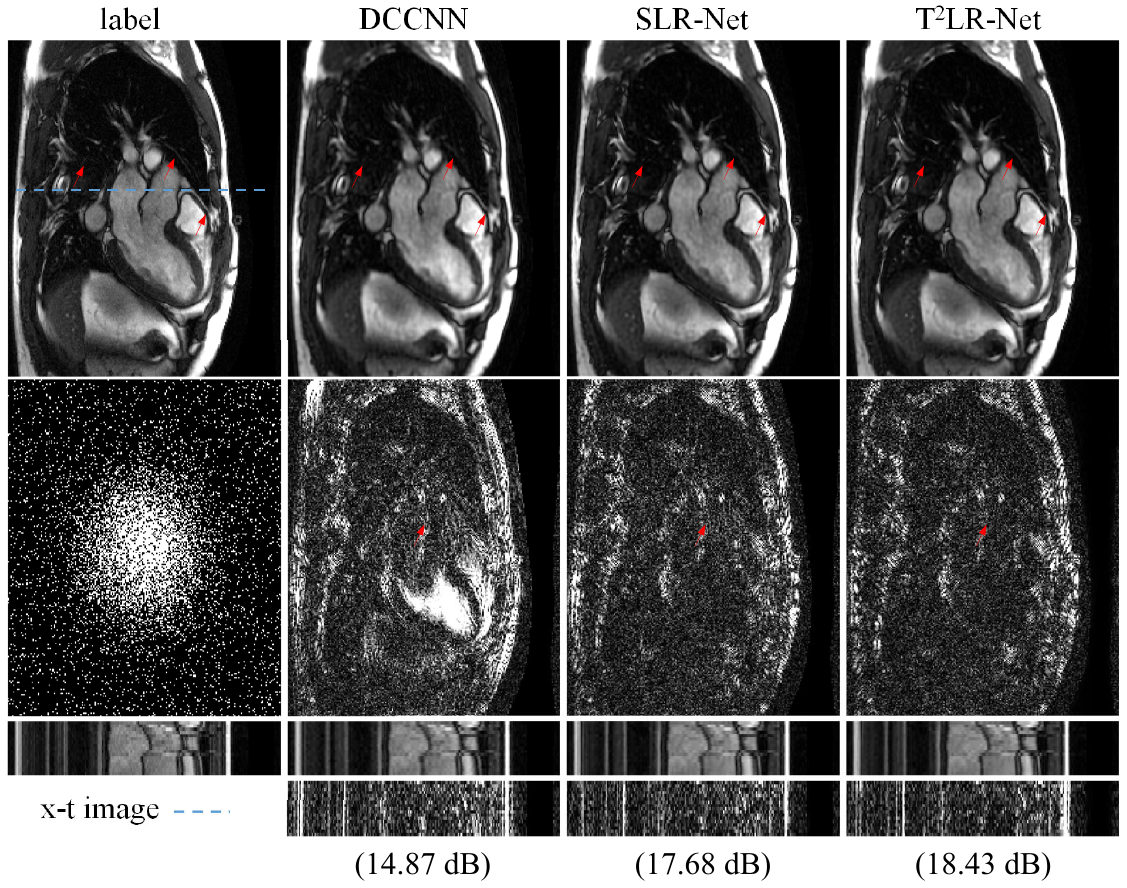}
  \caption{The OCMR reconstruction results of different methods under the Vds pattern with 10 acceleration rate in multi-coil scenario.}
  \label{multicoil}
\end{figure}

We further evaluate the proposed T$^2$LR-Net in the multi-coil scenario on the OCMR dataset. The quantitative metrics of our proposed T$^2$LR-Net and two SOTA unrolling networks are listed in Table \ref{tab:multicoil}. Two sampling cases are considered in this scenario, i.e., Radial sampling pattern with 16 lines and Vds pattern with 10 acceleration rate. The results show that T$^2$LR-Net achieves the highest SNR and SSIM in both sampling cases. The SNR of the results by T$^2$LR-Net is improved by 3.3 dB and 1.1 dB compared with SLR-Net in the Radial and Vds sampling cases, respectively. The SSIM of the results by T$^2$LR-Net is also improved by 1.7\% and 0.5\% compared with SLR-Net in the Radial and Vds sampling cases, respectively. The reconstruction results under Vds pattern are shown in Figure \ref{multicoil}. Our proposed T$^2$LR-Net produces sharper edges and clearer texture details than the other two methods. These results demonstrate that the proposed T$^2$LR-Net can effectively improve the reconstruction performance in the multi-coil scenario. It is worth noting that in the multi-coil scenario, the reconstruction performance of all methods degrades compared to the single-coil scenario. This is due to the imperfect estimation of coil sensitivity maps by ESPIRiT, which increases the difficulty of achieving accurate reconstructions \cite{muckley2021results}.

\begin{table}[htbp]
  \centering
  
  \caption{The quantitative metrics of different methods under two sampling cases in multi-coil scenario on the OCMR dataset (Mean ± Standard deviation).}
    \resizebox{1\linewidth}{!}{\begin{tabular}{cc|ccc}
    \toprule
          &       & \textbf{DCCNN} & \textbf{SLR-Net} & \textbf{T$^2$LR-Net} \\
    \midrule
    \multirow{2}[2]{*}{\textbf{Radial-16}} & SNR   & 15.258±1.131 & 17.676±1.840 & \textbf{20.941±2.271} \\
          & SSIM  & 0.938±0.027 & 0.952±0.015 & \textbf{0.969±0.013} \\
    \midrule
    \multirow{2}[2]{*}{\textbf{Vds-10}} & SNR   & 13.509±1.146 & 15.456±1.671 & \textbf{16.597±1.580} \\
          & SSIM  & 0.925±0.025 & 0.942±0.019 & \textbf{0.947±0.019} \\
    \bottomrule
    \end{tabular}}%
  \label{tab:multicoil}%
\end{table}%

% \vspace{-5pt}
\subsection{Prospective Experiments}

\begin{figure}[htbp]
  \centering
  \includegraphics[scale=0.55]{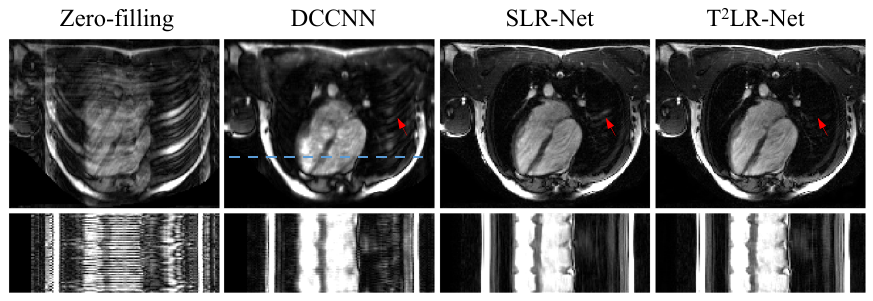}
  \caption{Prospective results of different methods. We use the 18-coil undersampled raw data with the VISTA sampling pattern of the OCMR dataset. The x-t images indicated by the blue dot line are shown in the second row.} 
  \label{prospect}
\end{figure}

We use an 18-coil real undersampled data from the OCMR dataset, acquired on a 1.5T Siemens Avanto scanner, with the VISTA \cite{ahmad2015vista} sampling pattern at 8-fold acceleration. To facilitate the prospective experiments, we retrain unrolling networks using the same sampling pattern and acceleration rate on OCMR dataset for 50 epochs. We show the reconstruction results of different methods in Figure \ref{prospect}. DCCNN produces blurry results and the reconstruction results of SLR-Net show some artifacts, while our proposed T$^2$LR-Net produces sharper edges and clearer texture details. It is noticed that the reconstruction performance of all networks seems to degrade compared to the retrospective experiments. This is because VISTA only undersamples the phase encoding lines while keeping the frequency encoding lines fully sampled. The incoherence between samples is less pronounced compared to pseudo-radial or Vds sampling, which poses challenges in reconstruction.

% \clearpage
\section{Discussion}
\label{discuss}
In this subsection, we discuss the efficacy of the transformed tensor low-rank prior and also study the necessity of the symmetric constraint on the CNN-learned transforms. All the networks in this section are trained using a pseudo-radial sampling mask with 16 lines for 50 epochs on the OCMR dataset. 

\subsection{The Efficacy of The Proposed Transformed Tensor Low-rank Prior: A Study of Two Important Indicators}
\label{subsec:u_lr}

\begin{figure}[htbp]
  \centering
  \includegraphics[scale=0.6]{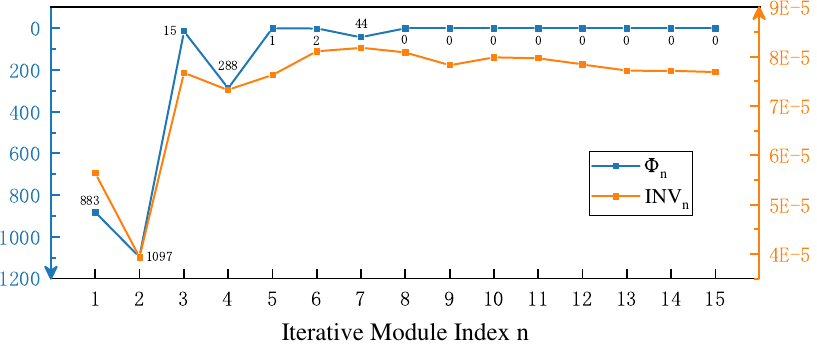}
  \caption{The curves of the indicators $\Phi_n$ and $\operatorname{INV}_n$, where the blue y-axis regarding $\Phi_n$ is reversed and the value of indicator $\Phi_n$ is labeled on the corresponding curve.}
  \label{inner}
\end{figure}

To evaluate the effectiveness of the transformed tensor low-rank (T$^2$LR) prior, we analyze two key indicators on the T$^2$LR-Net. The first indicator, $\Phi_n$, represents the number of singular values that fall below the thresholds of the $n$-th iteration module. It is defined as follows:
\begin{equation}
  \Phi_n = \sum_{i=1}^{n_t} \sum_{m=1}^{\operatorname{min}{n_x,n_y}} \mathbb{I}(\sigma_{n,i}^m < \tau_{n,i}),
\end{equation}
where $\sigma_{n,i}^m$ is the $m$-th singular value of the $i$-th frontal slice $\mathsf{T}_n(\tX_{n-1} + \t{L}_{n-1})^{(i)}$ of the $n$-th iteration module, $\tau_{n,i}$ is the corresponding threshold, and $\mathbb{I}$ is the indicator function. 

The second indicator, $\operatorname{INV}_n$, measures the invertibility of the two transforms learned by the CNNs in the $n$-th iteration module, and it is expressed as 
\begin{equation}
  \operatorname{INV}_n = \Vert \widetilde{\mathsf{T}_n} \circ \mathsf{T}_n(\tX_{n-1}) - \tX_{n-1} \Vert_F.
\end{equation}
Note that we use this indicator as a relaxed measurement of the Hermitian symmetry of the transforms. 

We evaluate these indicators using a test image from the OCMR dataset and plot their values across iteration modules in Figure \ref{inner}. Note that the blue y-axis regarding $\Phi_n$ in the figure is reversed. The value of indicator $\Phi_n$ is labeled on the corresponding curve. The two curves exhibit a remarkable degree of similarity. The results show that the T$^2$LR prior is given more emphasis in the first four iteration modules because of the significantly high $\Phi_n$ values. In contrast, $\operatorname{INV}_n$ values are relatively small, which is consistent with the Hermitian symmetric requirement of the transforms $\widetilde{\mathsf{T}_n}$ and $\mathsf{T}_n$ in the Equation \eqref{eq_zblock}. In the last eight iteration modules, the low-rank prior is no longer used since $\Phi_n$ becomes zero, which renders the singular value thresholding (SVT) operator in \eqref{eq_zblock} invalid. Consequently, the transformed tensor low-rank block degenerates into a cascade of two CNNs that learn implicit and complex priors beyond low rank to enhance the reconstruction performance. The increasing $\operatorname{INV}_n$ values imply that the symmetric relation between the two CNN-learned transforms is broken. The T$^2$LR-Net exploits the powerful learning capability of CNNs in the last iteration blocks. The iteration modules in the middle combine information from both the low-rank prior and the CNN-learned implicit prior to achieve the reconstruction. 

\begin{figure}[htbp]
  \centering
  \includegraphics[scale=0.5]{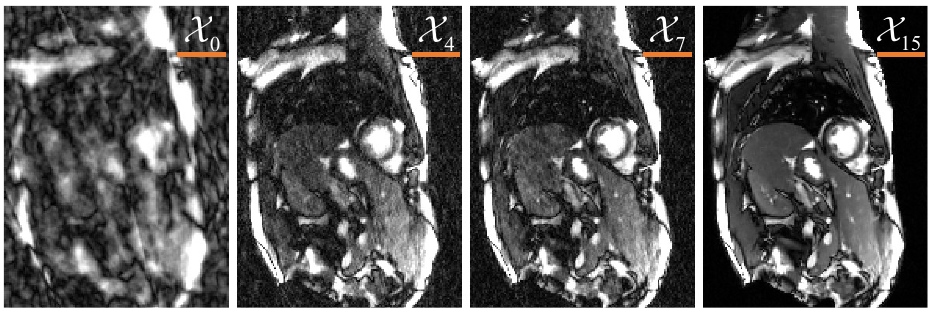}
  \caption{The reconstruction results from the iteration modules, where $\tX_0$ is the input aliased image and the other $\tX_i$ related to (\ref{iter_alg2}) is the result of the $i$th iteration module.}
  \label{inner2}
\end{figure}

We also display the reconstruction results from the iteration modules in Fig.\ref{inner2}, where $\tX_0$ is the input aliased image, and $\tX_i$ related to (\ref{iter_alg2}) is the output of the $i$th iteration module. The results reveal that the first four iteration modules effectively remove aliasing by utilizing the transformed tensor low-rank. The middle modules further preserve the edges and texture details, while the last eight modules adaptively suppress noise and smooth the tissues through the flexible learning CNN.

In summary, our proposed T$^2$LR-Net not only utilizes the transformed tensor low-rank prior but also incorporates the CNN-learned flexible and implicit prior. This is achieved by adaptively changing $\Phi_n$. The balance between the transformed low-rank and the implicit CNN-learned deep image prior is adjusted through the training process and adapted to the dataset.

% \vspace{-5pt}
\subsection{The Necessity of the Symmetric Constraint on the CNN-learned Transforms}
\label{subsubsection:loss_func}

\begin{figure}[htbp]
  \centering
  \includegraphics[scale=0.4]{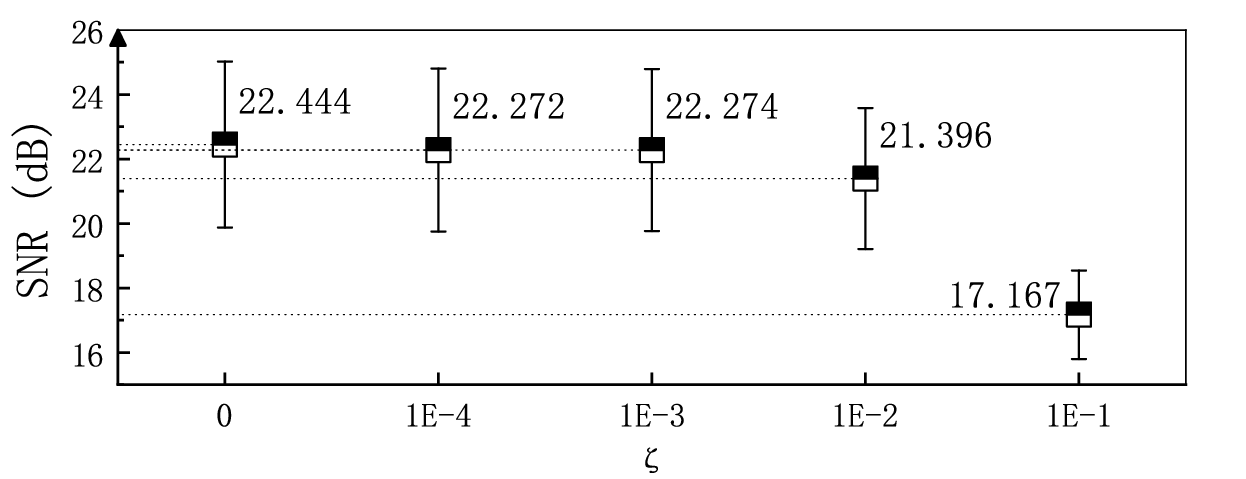}
  \caption{Effect of hyperparameter $\zeta$ of the reformulated loss function. The mean SNR and the standard deviation are plotted.}
  \label{fig:zeta}
\end{figure}

In line with the approaches of SLR-Net \cite{ref_slrnet} and ISTA-Net \cite{ref_ISTANET}, we explore the impact of incorporating an inverse constraint (as a relaxed version of the unitary constraint) for the two CNNs in the loss function. The combined loss function is reformulated as:
\begin{equation}
  \label{loss_func}
  \begin{aligned}
      Loss = \sum_{(\bar{\tX}, \mathbf{b}) \in \Omega}&\Vert \bar{\tX} - f_{cnn}(\mathbf{b}|\theta) \Vert_F^2 + \\%
      \zeta \sum_{n=1}^{N}&\Vert \widetilde{\mathsf{T}_n} \circ \mathsf{T}_n(\tX_{n-1}) - \tX_{n-1} \Vert_F^2.
  \end{aligned}
\end{equation}

We retrain the network with varying $\zeta$ values and report the reconstruction SNRs on the OCMR test dataset. The T$^2$LR-Net that we proposed is a specific case where $\zeta=0$. The results, presented in Figure \ref{fig:zeta}, show that the average SNR increases from 17.17 dB to 22.27 dB as $\zeta$ decreases from 1E-1 to 1E-4. The best average SNR of 22.44 dB is achieved with $\zeta=0$.

We observe that despite adopting the relaxed invertibility constraint instead of Hermitian symmetry, the performance worsens with an increasing weight of the invertibility constraint in the loss function. As discussed in Section \ref{subsec:u_lr}, if $\mathsf{T}_n \circ \widetilde{\mathsf{T}_n}(\tX) = \tX$ strictly holds, the CNNs in the iteration modules where $\Phi_n$ equals 0 (i.e., the ones where the SVT operation is invalid) will vanish. Thus, the deep implicit prior is not learned, and the reconstruction performance is limited. Based on this observation, we can conclude that there is no need to impose the symmetry constraint on the CNN-learned transforms.

\subsection{Effect of the number of iteration modules}

We retrain the networks with 5, 10, 15 and 20 iteration modules and compare the reconstruction results on the test dataset of OCMR to investigate the effect of the number of iteration modules. The results are shown in Tab.\ref{Neffect}. From the results, it is observed that the improvement in SNR slows down after more than 15 iterations. Therefore, considering the trade-off between efficiency and accuracy, we chose 15 iteration modules. However, it is important to note that without considering efficiency, a higher number of iteration modules would lead to higher reconstruction accuracy.

\begin{table}[htbp]
    \centering
    \caption{Effect of the number of iteration modules. The SNR values are shown in the form of Mean ± Standard deviation.}\label{Neffect}
    \setlength{\tabcolsep}{1.5mm}
    \resizebox{1\linewidth}{!}{\begin{tabular}{ccccc}
    \toprule
    \textbf{num} & 5     & 10    & 15    & 20 \\
    \midrule
    \textbf{SNR}(dB) & 19.38 ± 1.78 & 21.70 ± 2.38 & 22.44 ± 2.58 & 22.71 ± 2.66 \\
    \bottomrule
    \end{tabular}}%}
\end{table}%

\subsection{Effect of different SVT strategies}

We compared the impact of different SVT strategies on our proposed unrolling network. We reiterate that our network utilizes a learnable SVT strategy, specifically applying soft thresholding to all singular values of the tensor with the learnable thresholds. Here, we contrast it with two other SVT strategies: fixed truncation with \emph{Soft/Hard} thresholding on the last 10\% of singular values. The comparison results are presented in Table \ref{tab:SVT}. From the results, it is evident that the soft thresholding strategy better leverages low-rank information, achieving superior reconstruction compared to the hard thresholding strategy. Additionally, the learned soft thresholding further enhances reconstruction performance by increasing the flexibility and better utilizing low-rank information.

\begin{table}[htbp]
  \centering
  \caption{Effect of different SVT strategies. The quantitative metrics are shown in the form of Mean ± Standard deviation.}\label{tab:SVT}
    \begin{tabular}{lcc}
    \toprule
          & \multicolumn{1}{l}{\textbf{SNR}} & \multicolumn{1}{l}{\textbf{SSIM}} \\
    \midrule
    \textbf{HardSVT-10\% thres} & \multicolumn{1}{l}{21.41±2.18} & \multicolumn{1}{l}{0.960±0.020} \\
    \textbf{SoftSVT-10\% thres} & \multicolumn{1}{l}{22.29±2.56} & \multicolumn{1}{l}{0.965±0.020} \\
    \textbf{SoftSVT-learned thres} & \textbf{22.44±2.60} & \textbf{0.968±0.020} \\
    \bottomrule
    \end{tabular}%
\end{table}%

% \vspace{-10pt}
\section{Conclusion and Future Work}
\label{conclude}

In this paper, we introduced T$^2$LR-Net, a novel unrolling reconstruction network that exploits the transformed tensor low-rank prior for dynamic MR imaging. By extending the conventional t-SVD to a transformed version based on an arbitrary unitary transform, we outlined the mathematical definition of the UTNN and proposed a novel dynamic MR reconstruction model regularized by UTNN. The ADMM-based iterative algorithm was adopted to ensure efficient model solutions. By unfolding the iterative steps of the ADMM algorithm, we proposed a deep unrolling network that comprised numerous iteration modules, each employing a CNN to extract and exploit the tensor low-rank representation within the learned transformed domain. Both retrospective and prospective experimental results illustrated that T$^2$LR-Net surpassed existing optimization-based methods and unrolling network-based methods, thus proving its superior performance. Furthermore, although the proposed learned transformed tensor low-rank prior achieved significantly improved reconstruction performance, the SVT operators in each iteration module might be nondifferentiable \cite{chen2021efficient} in some special conditions, such as the matrix to be solved containing repeated singular values. It probably leads to the abortion of the training. In the future, we will explore the differentiable SVT operators to address this limitation.

% \vspace{-5pt}
\hypertarget{appendix}{\appendix}
\section*{Appendix}
\section{Derivation of UTNN}
\label{appd:ttnn}
\setcounter{equation}{0}
\renewcommand{\theequation}{A.\arabic{equation}}
According to the derivation in \cite{ref_tnn1}, the transformed tensor spectral norm w.r.t. $\mathsf{T}$, termed $\|\tX\|_\mathsf{T}$, induced by the Frobenius-normed operator norm, is defined by the matrix spectral norm of $\overline{\tX_\mathsf{T}}$, i.e., 
\begin{equation}
  \label{Snorm}
  \|\tX\|_\mathsf{T} = \|\overline{\tX_\mathsf{T}}\|.
\end{equation}
Then, according to the truth that the nuclear norm is the dual norm of the spectral norm, we define the UTNN as the dual norm of the transformed tensor spectral norm \cite{ref_tnn1}. For any $\mathcal{B} \in \mathbb{C}^{n_1 \times n_2 \times n_3}$, by \eqref{inner_prod} and \eqref{Snorm}, we have
\begin{align}
  \|\tX\|_{\mathsf{T}*} &= \mathop{\operatorname{sup}}_{\|\mathcal{B}\|_\mathsf{T} \leq 1}<\tX, \mathcal{B}> \\
  \label{trans_inner_prod}
  &= \mathop{\operatorname{sup}}_{\|\overline{\mathcal{B}_\mathsf{T}}\| \leq 1}<\overline{\mathcal{X}_\mathsf{T}}, \overline{\mathcal{B}_\mathsf{T}}> \\
  \label{relax_inner_prod}
  &\leq \mathop{\operatorname{sup}}_{\|\widetilde{\mathcal{B}}\| \leq 1}<\overline{\mathcal{X}_\mathsf{T}}, \widetilde{\mathcal{B}}> \\
  \label{nn_inner_prod}
  &= \|\overline{\mathcal{X}_\mathsf{T}}\|_*,
\end{align}
where equation \eqref{trans_inner_prod} is from \eqref{inner_prod} and \eqref{Snorm}, equation \eqref{relax_inner_prod} is because we relax the block diagonal matrix $\overline{\mathcal{B}_\mathsf{T}}$ into an arbitrary matrix $\widetilde{\mathcal{B}} \in \mathbb{C}^{n_1 \times n_2 \times n_3}$, equation \eqref{nn_inner_prod} holds due to that the matrix nuclear norm is the dual norm of the matrix spectral norm. Then, we show the equality \eqref{relax_inner_prod} holds and thus $\|\tX\|_{\mathsf{T}*} = \|\overline{\t{X}_\mathsf{T}}\|_*$. If $\mathcal{X} = \mathcal{U} \tmul \t{S} \tmul \t{V}^H$, we find an $\mathcal{B} = \t{U} \tmul \t{V}^H$, then we have,
\begin{align}
  <\tX, \mathcal{B}> &= <\mathcal{U} \tmul \t{S} \tmul \t{V}^H, \t{U} \tmul \t{V}^H>=  \\
  <\overline{\tX_\mathsf{T}}, \overline{\mathcal{B}_\mathsf{T}}> 
  \label{eq24}
  &= <\overline{\mathcal{U}_\mathsf{T}} \times \overline{\t{S}_\mathsf{T}} \times \overline{\t{V}_\mathsf{T}}^H, \overline{\t{U}_\mathsf{T}} \times \overline{\t{V}_\mathsf{T}}^H> \\
  \label{eq25}
  &= \operatorname{Tr}(\overline{\t{S}_\mathsf{T}})\\
  \label{eq26}
  &= \|\overline{\mathcal{X}_\mathsf{T}}\|_*,
\end{align}
where equation \eqref{eq24} is from \eqref{tprod_2} \eqref{eq_usv2}, and equation \eqref{eq25} is from the unitary property of $\t{U}$ and $\t{V}$, \eqref{eq26} holds due to the fact that the matrix nuclear norm is calculated by the sum of singular values. Thus, we have the definition of the UTNN (Definition.\ref{def:ttsvd}).

Please note that the matrix nuclear norm $\|\overline{\t{X}_\mathsf{T}}\|_*$ is the convex envelope of the matrix rank $\operatorname{rank}(\overline{\t{X}_\mathsf{T}})$ within the set $\{\overline{\t{X}_\mathsf{T}}|\|\overline{\t{X}_\mathsf{T}}\| \leq 1\}$. Thus, from \eqref{sum_rank}, \eqref{TTNN} and \eqref{Snorm}, the UTNN $\|\tX\|_{\mathsf{T}*}$ is the convex envelope of the transformed tensor sum rank $\operatorname{rank}_{sum}(\tX)$ on a unit ball of transformed tensor spectral norm $\{\tX \in \mathbb{C}^{n_1 \times n_2 \times n_3}|\|\tX\|_\mathsf{T} \leq 1\}$.

% \vspace{-10pt}
\section{Derivation of Transformed Tensor Singular Value Thresholding}
\label{appd:ttsvt}
\setcounter{equation}{0}
\renewcommand{\theequation}{B.\arabic{equation}}
Based on the definition of UTNN, the transformed tensor singular value thresholding ($\mathsf{T}$-TSVT) \cite{ref_tnn1,ref_ttnn} is derived as follows.

For any $\tau > 0$ and $\t{Y} = \mathcal{U} \tmul \mathcal{S} \tmul \mathcal{V}^H \in \mathbb{C}^{n_1 \times n_2 \times n_3}$, the closed solution of the following optimization problem is given by,
\begin{equation}
  \label{eq:tsvt}
  \mathsf{T}\mathrm{-TSVT}_{\tau}(\t{Y}) = \arg\min_{\tX} \tau \Vert \tX \Vert_{\mathsf{T}*}+\frac 12 \Vert \tX-\t{Y} \Vert_F^2,
\end{equation}
where ${\tau}$ becomes the threshold of $\mathsf{T}$-TSVT operator.

From \eqref{TTNN} and \eqref{Fnorm}, \eqref{eq:tsvt} can be converted into the following matrix optimization problem,
\begin{equation}
  \mathsf{T}^H\circ\operatorname{fold}\left(\mathop{\arg\min}_{\overline{\t{X}_\mathsf{T}}} \tau \|\overline{\t{X}_\mathsf{T}}\|_*+\frac 12 \Vert \overline{\t{X}_\mathsf{T}}-\overline{\t{Y}_\mathsf{T}} \Vert_F^2\right),
\end{equation}
where $\overline{\t{Y}_\mathsf{T}} = \overline{\mathcal{U}_\mathsf{T}} \times \overline{\t{S}_\mathsf{T}} \times \overline{\t{V}_\mathsf{T}}^H$, and the closed solution of the inside matrix optimization problem is given by the matrix SVT operator \cite{cai2010singular} with the threshold $\tau$,
\begin{equation}
  \operatorname{SVT}_{\tau}(\overline{\t{Y}_\mathsf{T}}) = \overline{\mathcal{U}_\mathsf{T}} \times \overline{\t{S}_{\mathsf{T}},\tau} \times \overline{\t{V}_\mathsf{T}}^H,
\end{equation}
where $\overline{\t{S}_{\mathsf{T}},\tau} = \max(\overline{\t{S}_{\mathsf{T}}}-\tau, 0)$.
Thus, we have
\begin{equation}
      \mathsf{T}\mathrm{-TSVT}_{\tau}(\t{Y}) = \mathsf{T}^H\circ\operatorname{fold}\circ\operatorname{SVT}_{\tau}\circ\mathsf{T}(\t{Y}).
\end{equation}
For clearness, we omit the $\operatorname{fold}$ operator, and thus obtain, 
\begin{equation}
  \label{ttsvt}
      \mathsf{T}\mathrm{-TSVT}_{\tau}(\t{Y}) = \mathsf{T}^H\circ\operatorname{SVT}_{\tau}\circ\mathsf{T}(\t{Y}).
\end{equation}

%%%%%%%%%%%%%%%%%%%%%%%%%%%%%%%%%%%%%%%%%%%%%%%%%%%%%%%%%%%%%%%%%%%%%%%%%
\section*{Acknowledgments}
This work is supported by National the Natural Science Foundation of China [grant number 62371167]; the Natural Science Foundation of Heilongjiang [grant number YQ2021F005].

\bibliographystyle{model1-num-names.bst}
\bibliography{refs}

% \section*{Supplementary Material}

% Supplementary material that may be helpful in the review process should
% be prepared and provided as a separate electronic file. That file can
% then be transformed into PDF format and submitted along with the
% manuscript and graphic files to the appropriate editorial office.

\end{document}